\documentclass[journal]{IEEEtran}
\usepackage{cite}

\ifCLASSINFOpdf
   \usepackage[pdftex]{graphicx}
  \graphicspath{{../pdf/}{../eps/}}
  \DeclareGraphicsExtensions{.pdf,.eps}
\else
\fi
\usepackage[cmex10]{amsmath}
\usepackage{amssymb}
\newtheorem{twr}{Theorem}[section]

\newtheorem{definition}[twr]{Definition}

\newtheorem{prz}[twr]{Example}
\newtheorem{obs}[twr]{Remark}
\newtheorem{uw}[twr]{Remark}

\usepackage{algorithmic}
\usepackage[tight,footnotesize]{subfigure}
\usepackage{url}

\begin{document}
%
\title{$k$-means Approach to the Karhunen-Lo\'eve Transform}
%
%
%

\author{Krzysztof~Misztal,
        Przemys\l aw~Spurek
        and~Jacek~Tabor
\thanks{K. Misztal, P. Spurek and J. Tabor are with Institute of Computer Science, Jagiellonian University, \L{}ojasiewicza 6, 30-348 Krak\'ow, Poland. e-mails: krzysztof.misztal@ii.uj.edu.pl; przemyslaw.spurek@ii.uj.edu.pl; jacek.tabor@ii.uj.edu.pl.}
\thanks{Manuscript received August 19, 2011.}}

\maketitle

\begin{abstract}
We present a simultaneous generalization of the well-known Karhunen-Lo\'eve (PCA) and $k$-means algorithms. The basic idea lies in approximating the data with $k$ affine subspaces
of a given dimension $n$. In the case $n=0$ we obtain the classical $k$-means, while for $k=1$ we obtain PCA algorithm.


We show that for some data exploration problems this method gives better result then either of the classical approaches.
\end{abstract}

\begin{IEEEkeywords}
Karhunen-Lo\'eve Transform, PCA, k-Means, optimization, compression, data compression, image compression.
\end{IEEEkeywords}

%
\IEEEpeerreviewmaketitle

\section{Introduction}

\IEEEPARstart{O}{ur} general problem concerns splitting of a given data-set $W$ into clusters with respect to their intrinsic dimensionality. The motivation to create such an algorithm is a desire to extract parts of data which can be easily described by a smaller number of parameters. More precisely, we want to find affine spaces $S_1,\ldots,S_k$ such that every element of $W$ belongs (with certain maximal error) to one of the spaces $S_1,\ldots,S_k$.

\begin{figure}[!h]
  \centering
	\subfigure[]{\label{fig:split1}\includegraphics[width=2.5in]{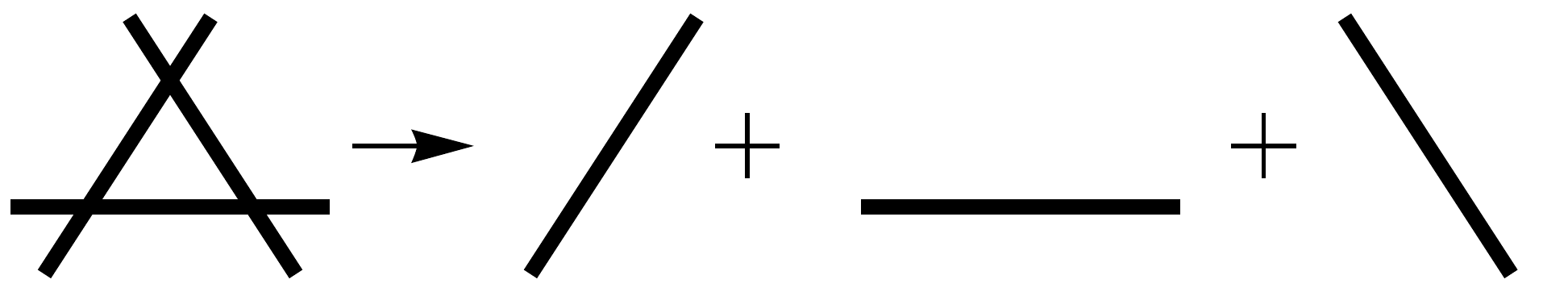}}\\
	\subfigure[]{\label{fig:split2}\includegraphics[width=2.5in]{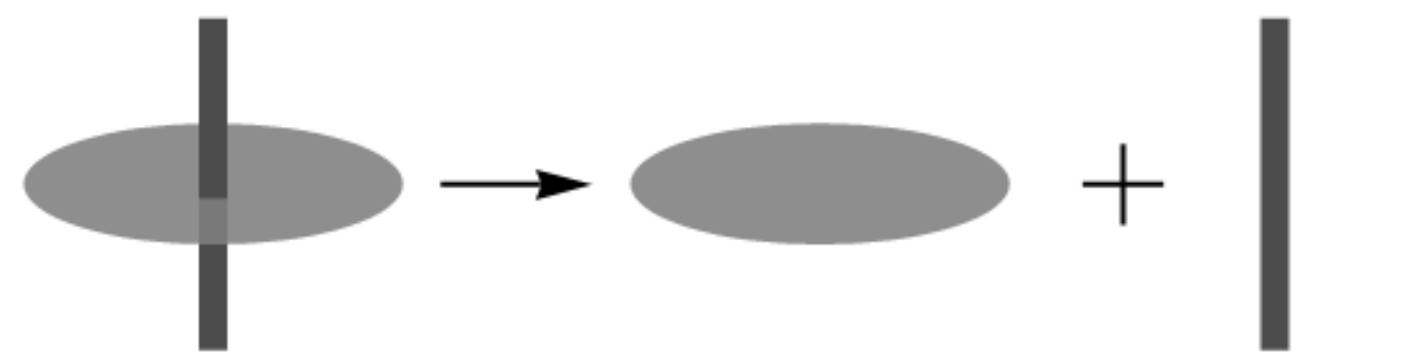}} 
	\caption{Our goal is to create algorithm which will interpret Fig. \ref{fig:split1} as three groups of one-dimensional points and  Fig. \ref{fig:split2} as two groups of one- and two-dimensional points.}
	\label{fig:split}
\end{figure}

To explain it graphically, let us consider following example. Figure \ref{fig:split1} represents three lines in the plane, while Figure \ref{fig:split2} a circle and an orthogonal line in the space. Our goal is to construct an algorithm that will split them into three lines and into a line and a circle.

We have constructed a simultaneous generalization of the $k$-means method \cite{Dubes} and the Karhunen-Lo\'eve transform (called also PCA -- Principle Component Analysis) \cite{Jol} -- we call it $(\omega,k)$-means. Instead of finding $k$ centers which best represent the data as in the classical $k$-means, we find $k$ nearest subspaces of a given fixed dimension $n$, where $\omega$
denotes the weight which takes part in measuring the distance. In analogy to the case of $k$-means, we obtain a version of the Voronoi diagram (for details see next section). In the simplest form our algorithm needs the number of clusters $k$ and the dimension $n$ (for $n=0$ we obtain the $k$-means while for $k=1$ we obtain the PCA).

\begin{figure}[!t]
  \centering
	\subfigure[]{\label{fig:klus-line-a}\includegraphics[width=1in]{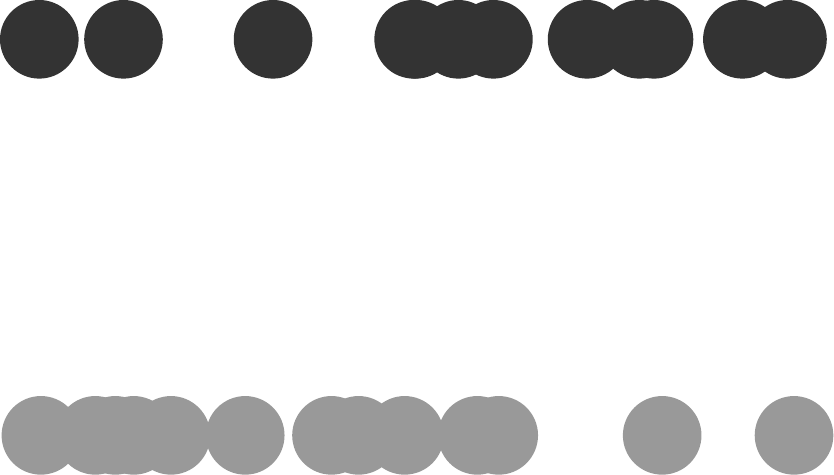}}\qquad\qquad 
	\subfigure[]{\label{fig:klus-line-b}\includegraphics[width=1in]{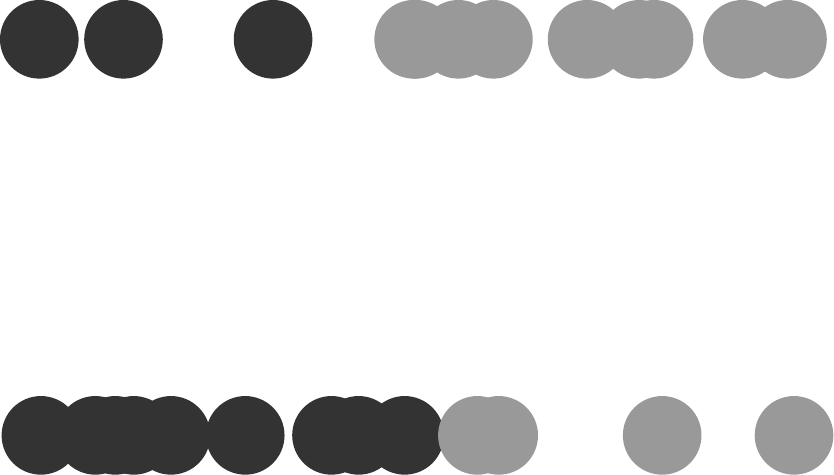}} 
	\caption{Example of clustering: Fig. \ref{fig:klus-line-a} for $k=2$ clusters which are 1--dimensional; Fig. \ref{fig:klus-line-b} for classical $k$-means with $k=2$ clusters.}
\label{fig:klus-line}
\end{figure}

To present our method, consider the points grouped along two parallel lines -- Figure \ref{fig:klus-line} presents the result of our program on the clustering of this set.

\begin{figure}[!t]
  \centering
	\includegraphics[width=3in]{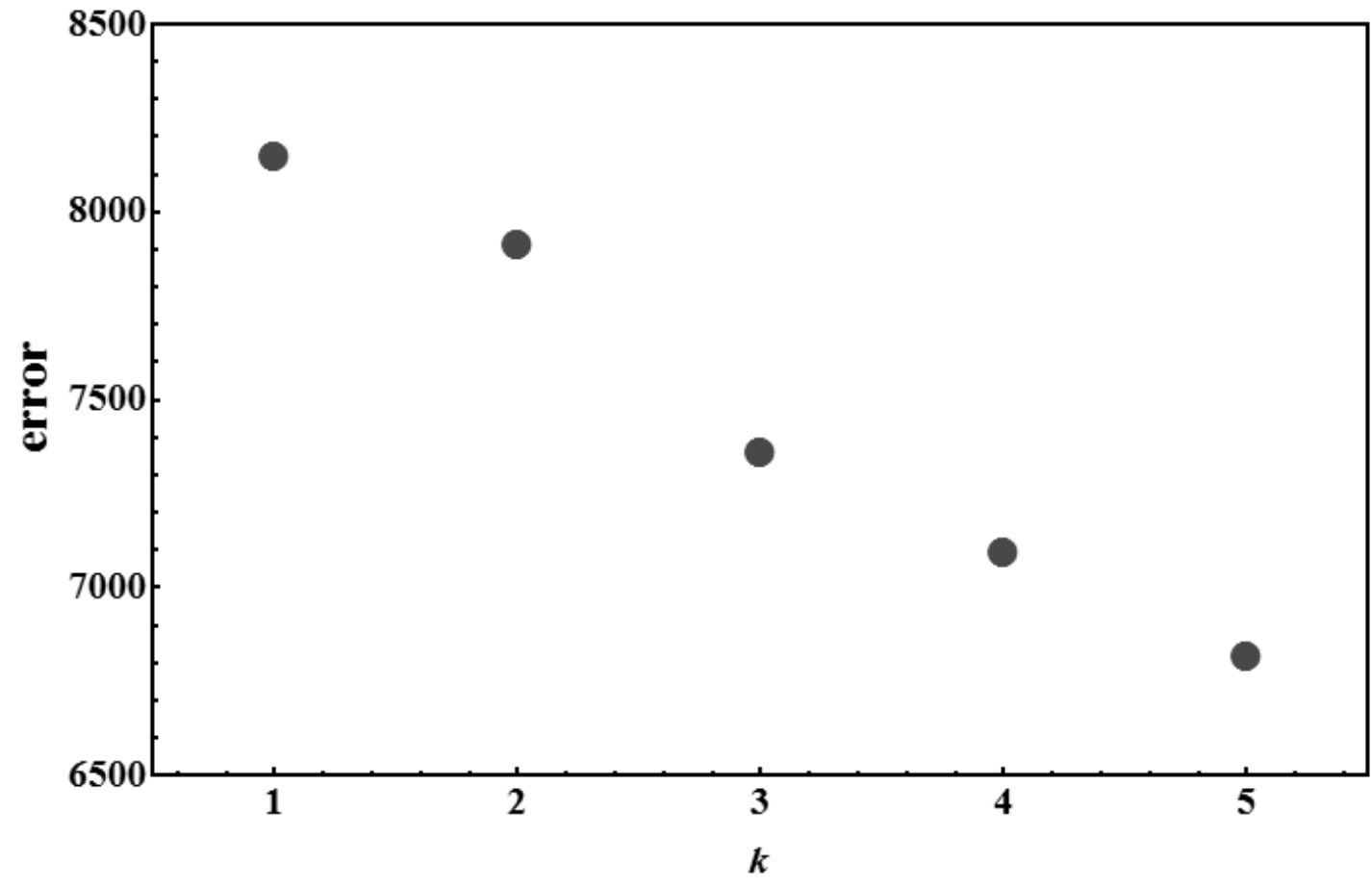}
	\caption{Error in image decompression as a function of number of clusters $k$ for $n=5$. }
	\label{fig:kerror}
\end{figure}

The approach can be clearly used in most standard applications of either the $k$-means or the Karhunen-Lo\'eve transform. In particular, (since it is a generalization of the Karhunen-Lo\'eve transform \cite{Salomon}) one of the possible natural applications of our method lies in the image compression. Figure \ref{fig:kerror} presents error\footnote{By error in image comparison we understand pixel by pixel image compare using standard Euclidean norm.} in image reconstruction of a classical Lena photo ($508 \times 508$ pixels) as a function of $k$. Observe that just by modifying the number of clusters from 1 to 3, which makes the minimal increase in the necessary memory, we decrease twice the level of error in the compression.

Except for image compression our method can by applied in various situations where the classical $k$-means or PCA where used, for example in: 
\begin{itemize}
	\item data mining -- we can detect important coordinates and subsets with similar properties;
	\item clustering -- our modification of $k$-means can detect different, high dimensional relation in data;
	\item image compression and image segmentation;
	\item pattern recognition --  thanks to detection of relation in data we can use it to assign data to defined before classes.
\end{itemize}

The basic idea of $(\omega,k)$-means algorithm can be described as follows:

\begin{center}
		\begin{algorithmic}
			\STATE {\bf choose}
			\STATE\hspace{\algorithmicindent} initial clusters distribution
			\REPEAT 
			\STATE \textit{apply} the Karhunen-Lo\'eve method for each cluster
			\STATE \textit{appoint} new clusters
			\UNTIL{decrease of "energy" is below given error}
		\end{algorithmic}

\end{center}

\section{Generalized Voronoi Diagram}

\begin{figure}[!t]
  \centering
	\subfigure[]{\label{fig:BDdef}\includegraphics[width=1.5in]{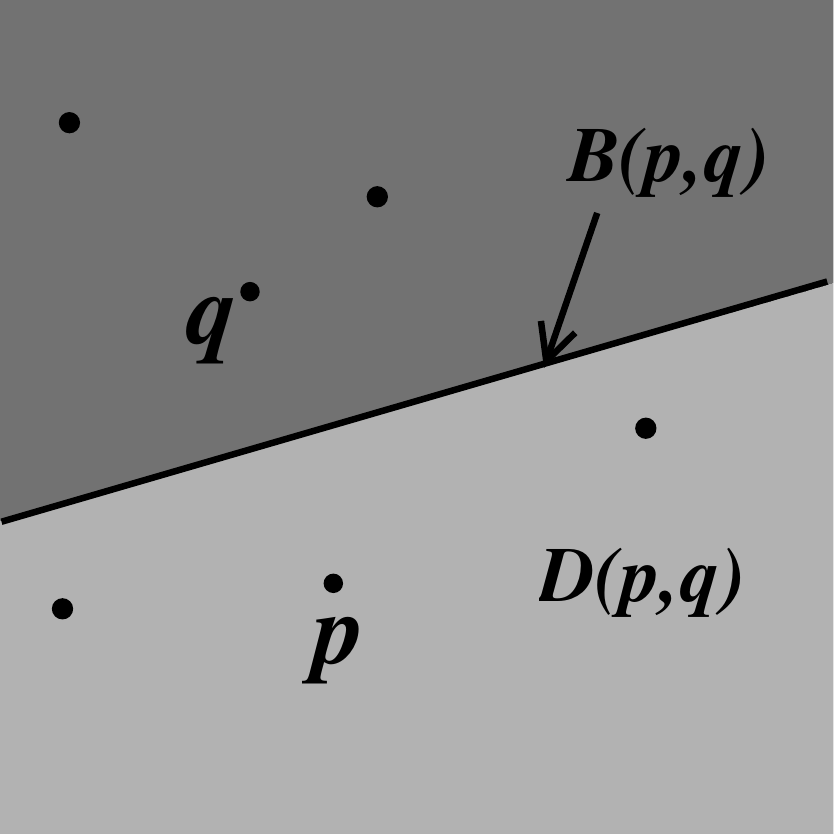}}\qquad 
	\subfigure[]{\label{fig:DSdef}\includegraphics[width=1.5in]{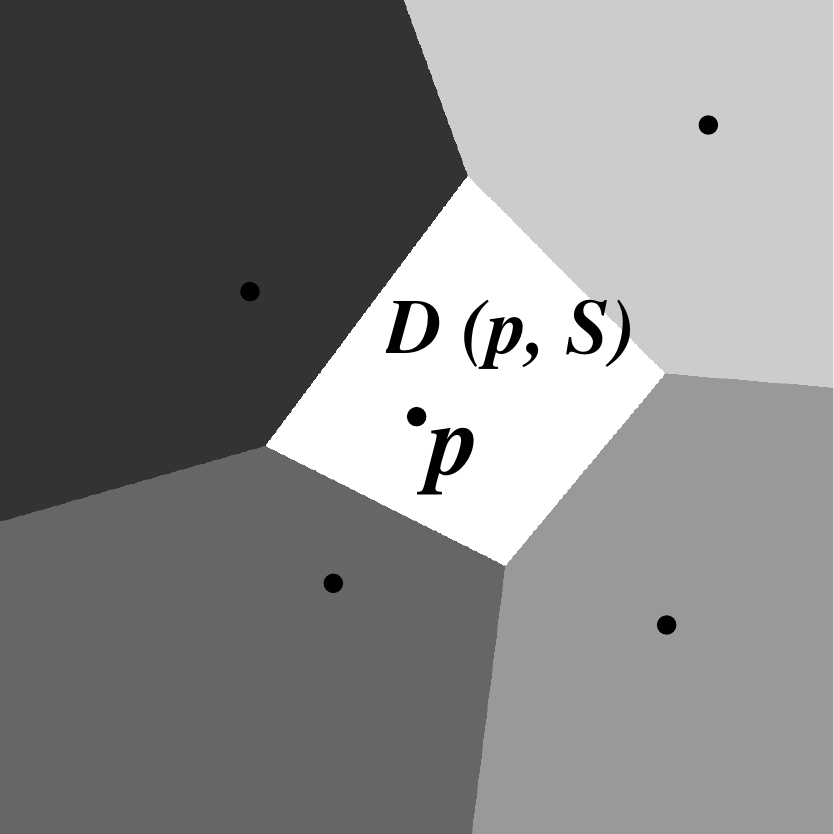}}  
	\caption{Graphical presentation of $B(p,q)$, $D(p,q)$ and $D(p,S)$ in $\mathbb{R}^2$.}
	\label{fig:BDSdef}
\end{figure}

The Voronoi diagram is one of the most useful data structures in computational geometry, with applications in many areas of science \cite{okabeVoronoi}. For the convenience of the reader and to establish the notation we shortly describe the classical version of the Voronoi diagram (for more details see \cite{Klein}). For $N \in \mathbb{N}$ consider $\mathbb{R}^{N}$  with the standard Euclidean distance and let $S$ be a finite set of $\mathbb{R}^N$. For $p, q \in S$ such, that $p \neq q$, let 
\begin{equation}\label{B}
	B(p,q) = \{ z \in \mathbb{R}^N \colon \| p-z \| = \| q-z \| \}, 
\end{equation}
\begin{equation}\label{D}
	D(p,q) = \{ z \in \mathbb{R}^N \colon \| p-z \| < \| q-z \| \}.
\end{equation}
Hyperplane $B(p,q)$ divides $\mathbb{R}^N$ into two sets, one containing points which are closer to point $p$ then $q$ ($D(p,q)$), and the second one containing points which are closer to point $q$ then $p$ ($D(q,p)$) -- see Figure \ref{fig:BDdef}. 

\begin{definition}[\cite{Klein}]
The set 
$$
	D(p,S):= \bigcap_{q \in S \colon q \neq p} D(p,q)
$$ 
of all points that are closer to $p$ than to any other element of $S$ is called the 
(open) Voronoi region of $p$ with respect to $S$. 
\end{definition}

For $N=2$ set $D(p,S)$ is the interior of a convex, possibly unbounded polygon (Figure \ref{fig:DSdef}).

The points on the contour of 
$D(p,S)$ are those that have more than one nearest neighbor in $S$, one of which  is $p$. 

\begin{definition}[\cite{Klein}]
The union
$$ 
	V(S) := \bigcup \partial D(p,S) 
$$
of all region boundaries is called the Voronoi diagram of $S$. 
\end{definition}

The common boundary of two Voronoi regions is a Voronoi edge. Two edges meet at a 
Voronoi vertex such a point has three or more nearest neighbors in the set $S$. 

Now we proceed to the description of our modification of the Voronoi diagram. We divide the space $\mathbb{R}^{N}$ with respect to affine subspaces of $\mathbb{R}^{N}$. 

\begin{definition}
For $n \leq N$ let
\setlength{\arraycolsep}{0.0em}
\begin{eqnarray*}
\mathrm{E}_{n} (\mathbb{R}^{N})&{}:={}& \{ ( \mathrm{v}_0,\ldots,\mathrm{v}_n) \in (\mathbb{R}^N)^{n+1} \mbox{ such that}\\
&&{}\:\mathrm{v}_{i}, \mathrm{v}_{j} \mbox{ are orthonormal for } i,j > 0, i \neq j \}.
\end{eqnarray*}
\setlength{\arraycolsep}{5pt}
\end{definition}
Thus $\mathrm{v}_{0}$ denotes a center of affine space we consider, while $\mathrm{v}_1,\ldots,\mathrm{v}_n$
is the orthonormal base of its "vector part". From the geometrical point of view the element $\mathrm{v}=(\mathrm{v}_0,\mathrm{v}_1,\ldots,\mathrm{v}_n) \in \mathrm{E}_n(\mathbb{R}^N)$ represents the affine space
$$
	\mathrm{v}_0+\mathrm{lin}(\mathrm{v}_1,\ldots,\mathrm{v}_n)=\mathrm{aff}(\mathrm{v}_0,\mathrm{v}_1,\ldots,\mathrm{v}_n).  
$$

We modify equations (\ref{B}) and (\ref{D}), by using distance between a point and affine subspace generated by linear independence vector instead of distance between points.

\begin{definition}
	Let $ n < N $ and let $\mathrm{v} \in \mathrm{E}_{n}(\mathbb{R}^N)  $, $ \omega=(\omega_0,\ldots,\omega_n ) \in [0,1]^{n+1}$ such that $\sum_{j=0}^n\limits \omega_j = 1$ be given. For $ x \in \mathbb{R}^{N} $ let
	\begin{equation}\label{DIST}
	\mathrm{DIST}_{\omega} (x;\mathrm{v}) := \left( \sum_{j=0}^{n} \omega_{j} \  \mathrm{dist}(x;\mathrm{aff}( \mathrm{v}_{0},\ldots,\mathrm{v}_{j}))^2 \right)^{1/2},
	\end{equation}
	where $\mathrm{dist}(x;V)$ denotes the distance of the point $x$ from the space $V$.
\end{definition}

In formula \eqref{DIST}, $\omega=(\omega_0,\ldots,\omega_n)$ is interpreted as vector of weights, where $\omega_k$ denotes the weight of the affine subspace of dimension $k$. It is easy to notice, that  $\mathrm{DIST}$ has following properties:
\begin{itemize}
	\item for $\mathrm{v} \in \mathrm{E}_n(\mathbb{R}^N)$ and $\omega=(0,\ldots,0,1) \in [0,1]^{n+1}$ we obtain that $\mathrm{DIST}_\omega(x;\mathrm{v})$ is a distance between the point $x$ and affine space $\mathrm{aff}(\mathrm{v})$;
	\item if $\mathrm{v}_{0}=0$ and $\omega=(0,\ldots,0,1)$ then $\mathrm{DIST_\omega}$ is a distance between point and linear space generated by $(\mathrm{v}_1,\ldots,\mathrm{v}_n)$;
	\item if $\omega=(1,0,\ldots,0)$ then $\mathrm{DIST}_\omega$ is the classical distance between $x$ and $\mathrm{v}_0$:
	$$ 
	\mathrm{DIST}_{\omega} ( x;\mathrm{v}) = \| x - \mathrm{v}_{0} \|.
	$$
\end{itemize}

\begin{obs} \label{rem:2.5}
Formula \eqref{DIST} can be computed as follows
\begin{align*}
	\left( \mathrm{DIST}_{\omega} (x;\mathrm{v}) \right)^2=&\sum_{j=0}^n \omega_j \left( \|x-\mathrm{v}_0\|^2-\sum_{i=1}^j \langle x-\mathrm{v}_0;\mathrm{v}_i\rangle^2 \right)\\
	=& \sum_{j=0}^n \omega_j \|x-\mathrm{v}_0\|^2-\sum_{j=0}^n\omega_j\sum_{i=1}^j  \langle x-\mathrm{v}_0;\mathrm{v}_i\rangle^2.
	\end{align*}
	To optimize calculations we define 
	\begin{align*}
	\bar{\mathrm{v}}_1 &= \langle x-\mathrm{v}_0;\mathrm{v}_1\rangle^2,\\
	\bar{\mathrm{v}}_j &=  \bar{\mathrm{v}}_{j-1}+\langle x-\mathrm{v}_0;\mathrm{v}_j\rangle^2,
\end{align*}
and since $\sum \omega_j = 1$ thus we simplify our computation to
\begin{align*}
	\left( \mathrm{DIST}_{\omega} (x;\mathrm{v}) \right)^2 =& \|x-\mathrm{v}_0\|^2-\sum_{j=0}^n\omega_j \bar{\mathrm{v}}_j.
\end{align*}
\end{obs}

Now we are ready to define our generalization of the Voronoi diagram. Let $ S$ be a finite subset of  $\mathrm{E}_{n}(\mathbb{R}^N) $ and $\omega \in [0,1]^{n+1}$, $\sum \omega_j = 1$, where $n \leq N$. For $\mathrm{p},\mathrm{q} \in S$ such, that $\mathrm{p} \neq \mathrm{q}$, let 
$$
	B_{\omega}(\mathrm{p},\mathrm{q}) := \{ z \in \mathbb{R}^N \colon \mathrm{DIST}_{\omega}( z ; \mathrm{p} ) = \mathrm{DIST}_{\omega} ( z ; \mathrm{q} ) \},
$$
$$
	D_{\omega}(\mathrm{p},\mathrm{q}) := \{ z \in \mathbb{R}^N \colon \mathrm{DIST}_{\omega} ( z ; \mathrm{p} ) < \mathrm{DIST}_{\omega} ( z ; \mathrm{q} ) \}.
$$
The set $B_{\omega}(\mathrm{p},\mathrm{q})$ divides the space $\mathbb{R}^N$ into two sets, first containing points which are closer to $\mathrm{p}$ then to $\mathrm{q}$ ($D_{\omega}(\mathrm{p},\mathrm{q})$) and second contain points which are closer to $\mathrm{q}$ then $\mathrm{p}$ ($D_{\omega}(\mathrm{q},\mathrm{p})$). 

\begin{definition}
	Let $n\in \mathbb{N}$, $n < N$ be fixed. Let $S$ be finite subset of $\mathrm{E}_{n}(\mathbb{R}^{N}) $ and $\omega \in [0,1]^{n+1} $, $\sum_{j=0}^n \omega_j = 1$ be given.
	For $\mathrm{p} \in S$ the set 
	$$
		D_{\omega}(\mathrm{p},S):= \bigcap_{ \mathrm{q} \in S \colon \mathrm{q} \neq \mathrm{p} } D_{\omega}( \mathrm{p}, \mathrm{q})
	$$ 
	of all points that are closer to $\mathrm{p}$ than to any other element of $S$ is called the (open) generalized Voronoi region of $\mathrm{p}$ with respect to $S$. 
\end{definition}

\begin{figure}[!t]
\centering
	\includegraphics[height=2in]{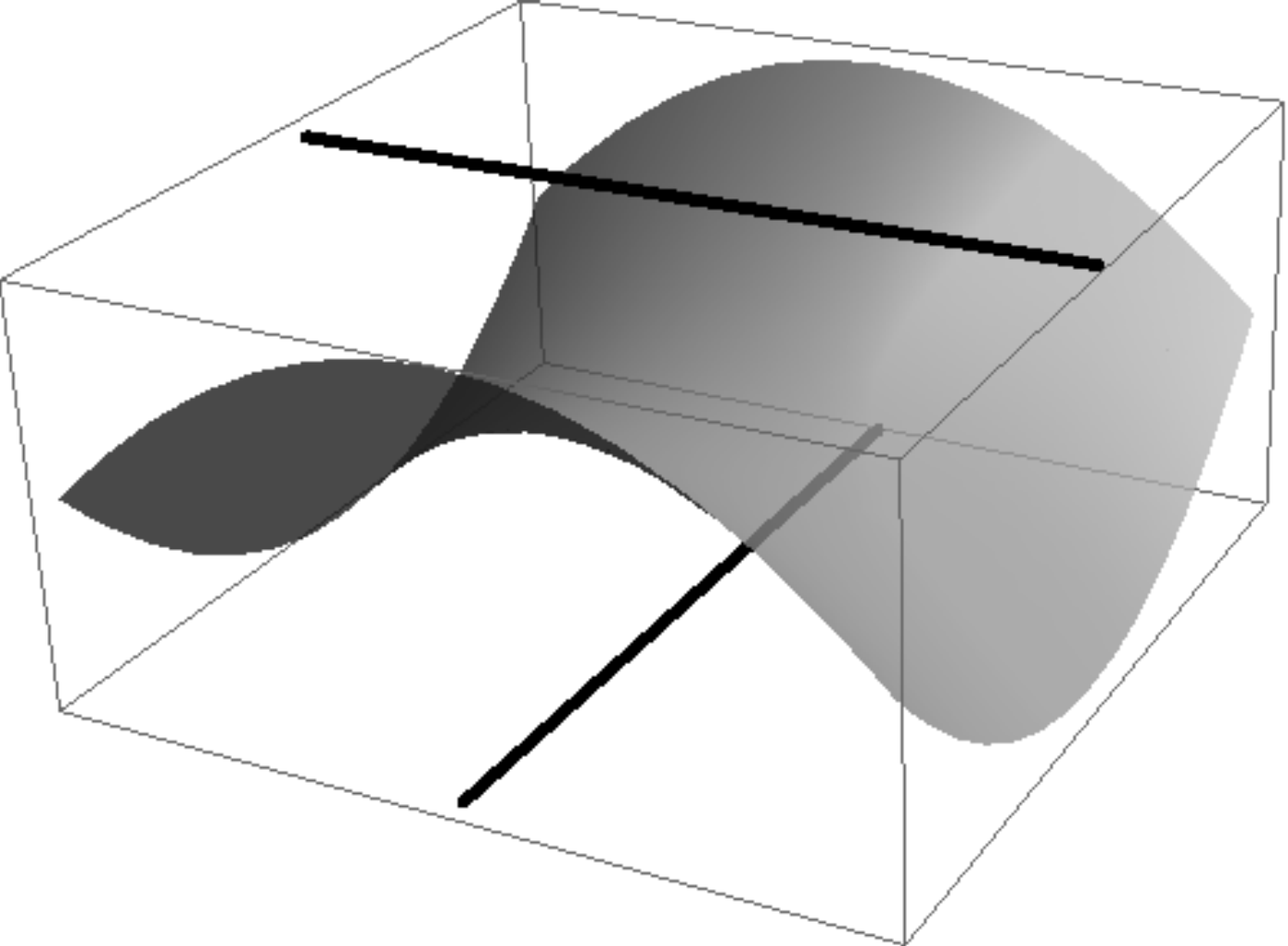}
	\caption{Generalized Voronoi diagram for $\omega=(0,1)$ and two lines.}
	\label{fig:vor3d}
\end{figure}

Applying this definition we obtain a new type of Voronoi diagram.
\begin{figure}[!t]
  \centering
	\subfigure[two lines]{\label{fig:diag2}\includegraphics[width=1in]{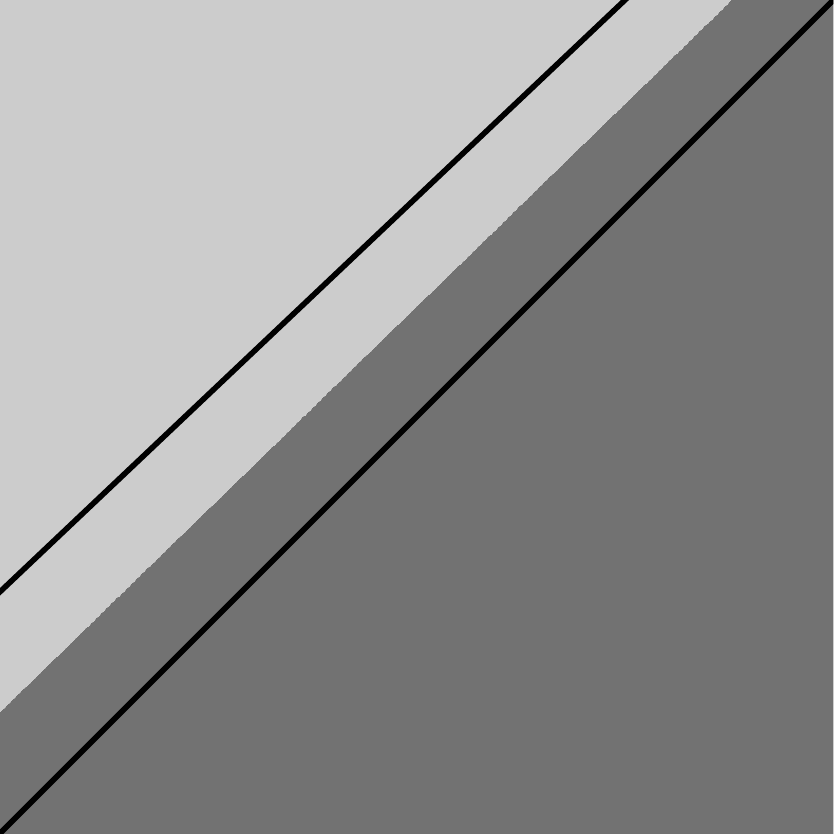}}\quad 
	\subfigure[three lines]{\label{fig:diag3}\includegraphics[width=1in]{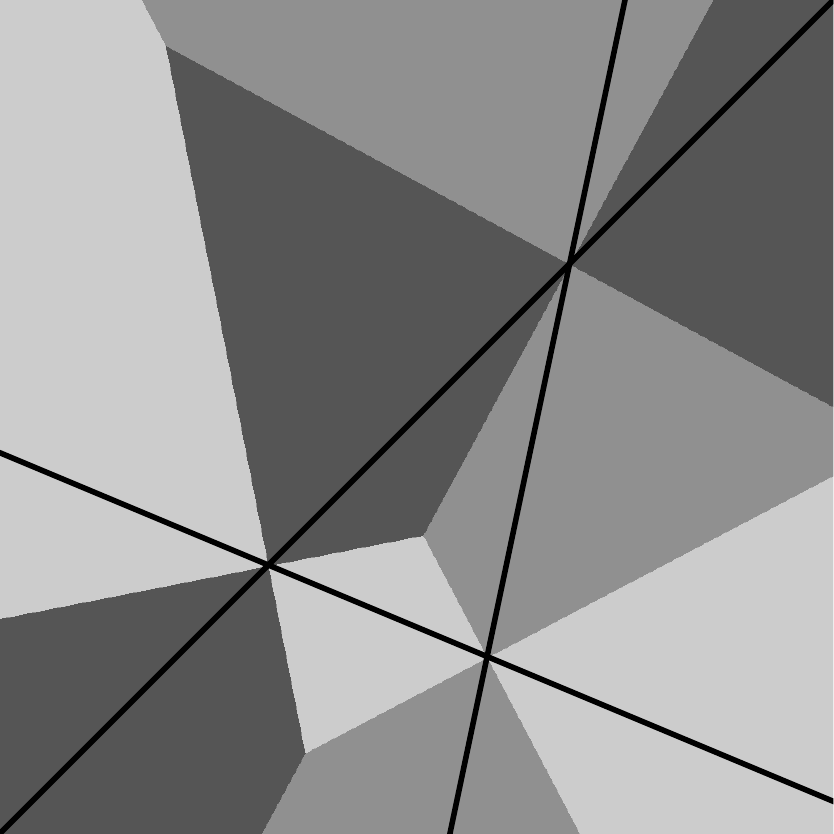}}\quad
	\subfigure[four lines]{\label{fig:diag4}\includegraphics[width=1in]{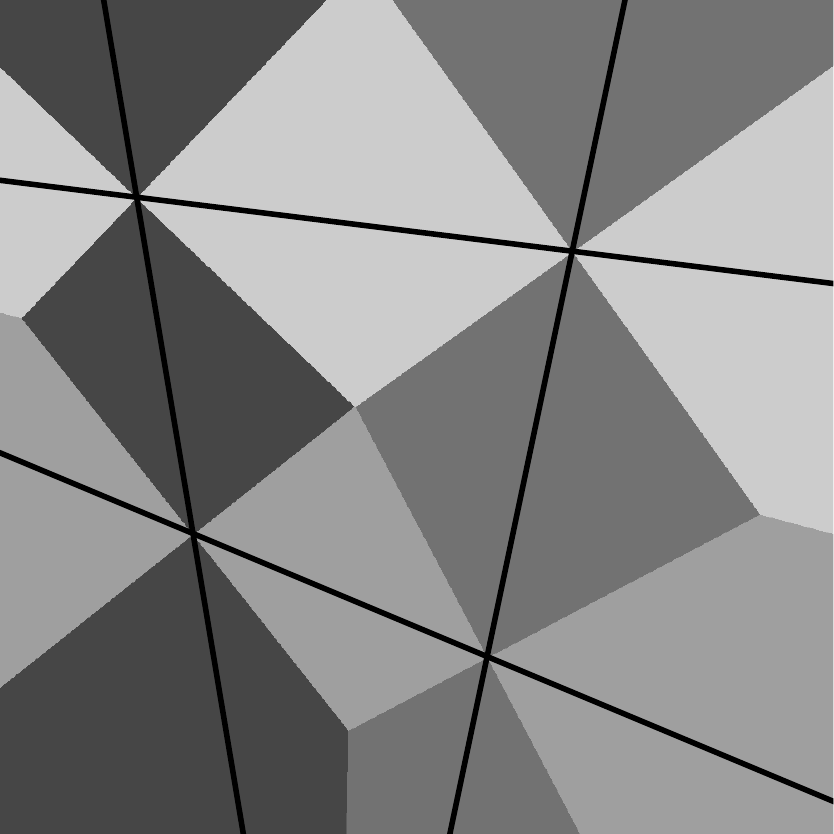}}\quad 
	\caption{Generalized Voronoi diagram for $\omega=(0,1)$ and two, three and four lines on plane.}
	\label{fig:vor2d}
\end{figure}
As we can see in Figure~\ref{fig:vor2d}, if $\omega=(0,1)$ we divide the plane into (not necessarily convex) polygons (similar situation to the classical Voronoi diagram). 
\begin{figure}[!t]
\centering
	\subfigure[]{\label{fig:wk100}\includegraphics[width=1in]{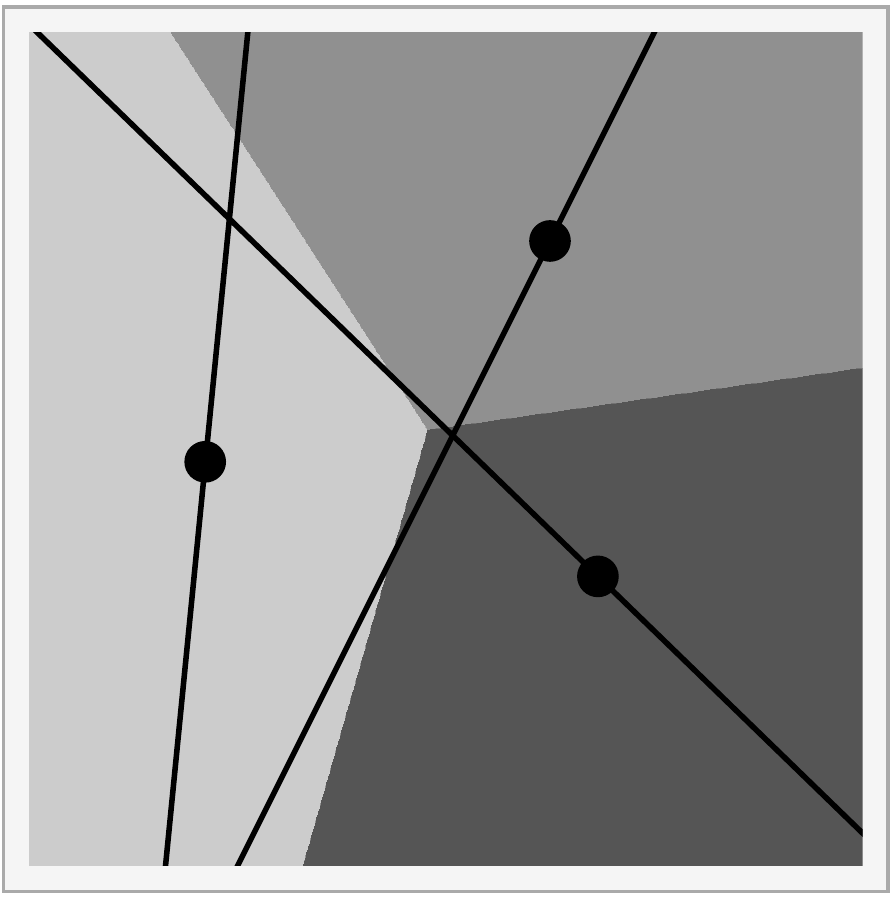}}\quad
	\subfigure[]{\label{fig:wk075}\includegraphics[width=1in]{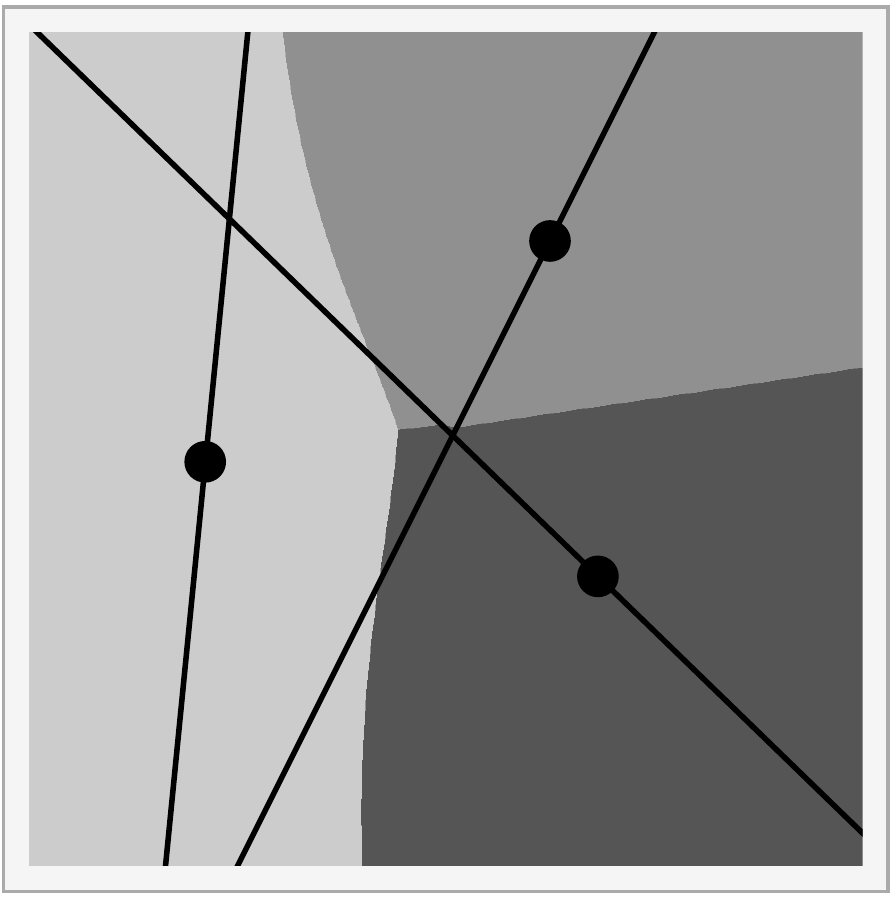}}\quad	
	\subfigure[]{\label{fig:wk050}\includegraphics[width=1in]{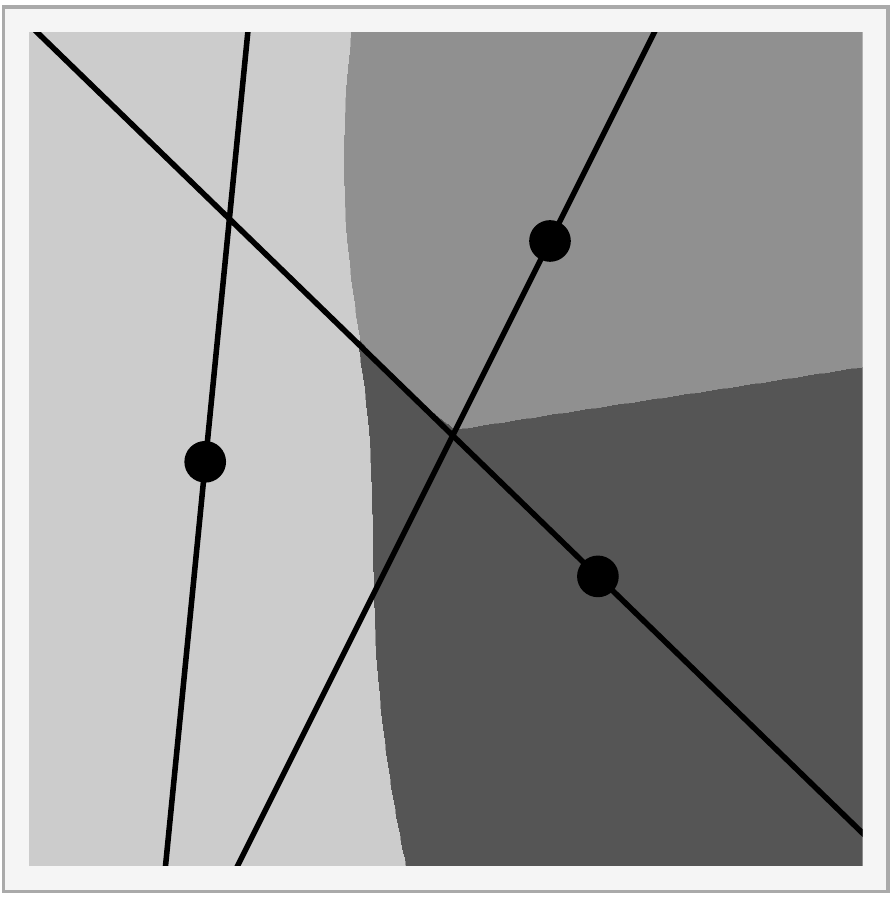}}\quad
	\subfigure[]{\label{fig:wk025}\includegraphics[width=1in]{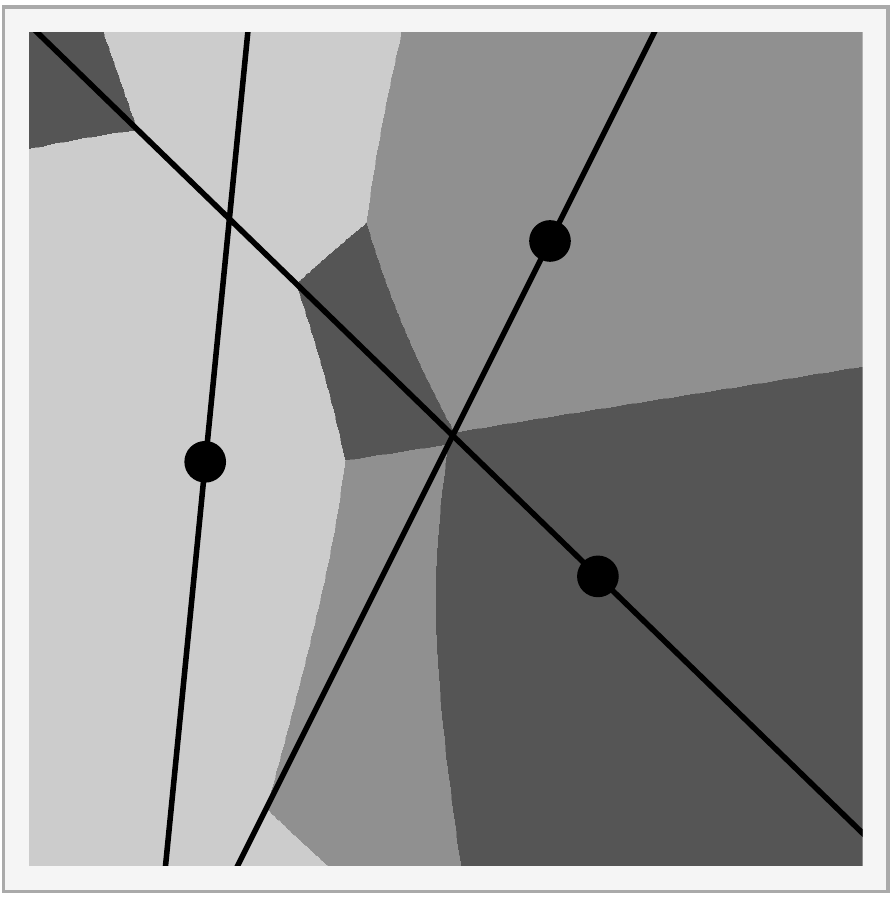}}\quad 
	\subfigure[]{\label{fig:wk000}\includegraphics[width=1in]{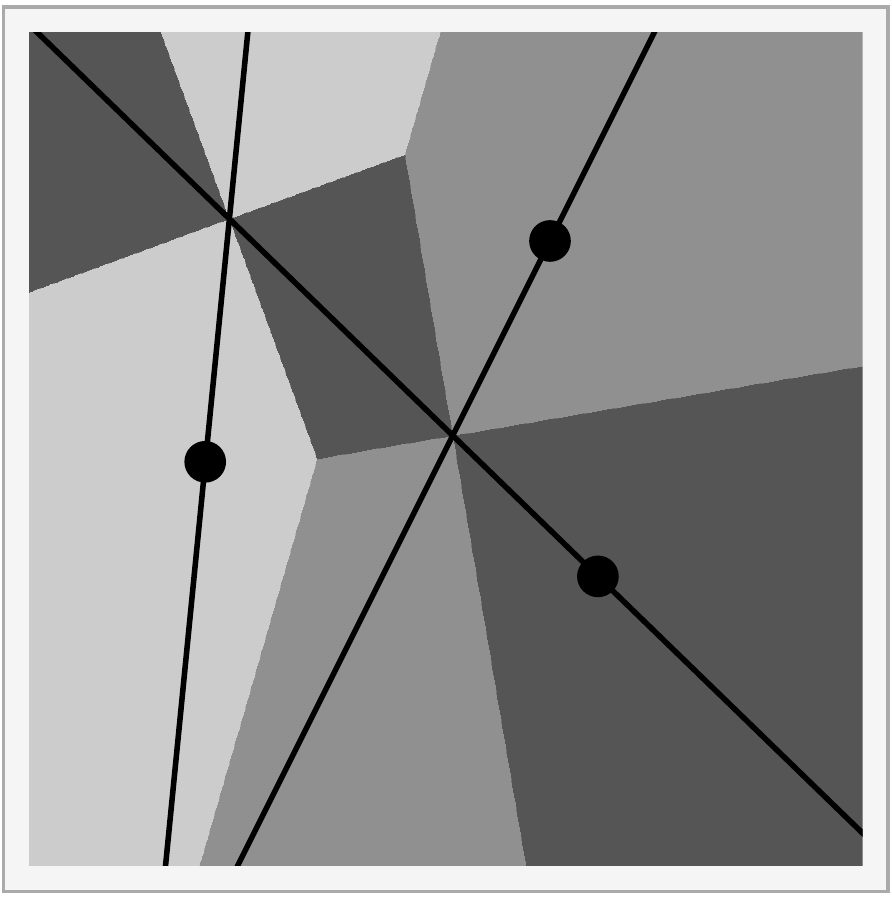}}
	\caption{Generalized Voronoi diagram for clustering of 3 clusters for different weight vectors Fig. \ref{fig:wk100}, $\omega=(1,0)$; Fig. \ref{fig:wk075}, $\omega=(\frac 3 4, \frac 1 4)$; Fig. \ref{fig:wk050}, $\omega=(\frac 1 2,\frac 1 2)$; Fig. \ref{fig:wk025}, $\omega=(\frac 1 4,\frac 3 4)$; Fig. \ref{fig:wk000}, $\omega=(0,1)$.}
	\label{fig:wk_all}
\end{figure}
Figure \ref{fig:wk_all} presents a generalized diagram on the plane for different weights changing from $\omega=(1,0)$ to $\omega=(0,1)$.
In general we obtain that the boundary sets usually are not polygons but
zeros of quadratic polynomials.
The same happens in $\mathbb{R}^3$ even for $\omega=(0,1)$  see the Figure \ref{fig:vor3d}, where we show points with equal distance from two lines.

\section{Generalization of the $k$-means method}

Clustering is a classical problem of the division of the set $ S \subset \mathbb{R}^{N}$ into separate clusters, or in other words, into sets showing given type of behavior. 
\subsection{$k$-means}
One of the most popular and basic method of clustering is the $k$-means algorithm.
By this approach we want to divide $S$ into $k$ clusters $ S_1, \ldots , S_k$ with minimal energy. 
For convenience of the reader and to establish the notation we shortly present the $k$-means algorithm.

For a cluster $S$ and $ r \in \mathbb{R}^N$ we define
$$
E(S,r) := \sum_{s \in S } \| s -  r \|^{2}.
$$
The function $\mathrm{E}(S,r)$ is often interpreted as an energy. 
We say that the point $  \overline{r}$ best ''describe" the set $S$ if the energy  is minimal, more precisely, if
$$
	\mathrm{E}(S, \overline{r}) = \inf _{r \in \mathbb{R}^N} \{ \mathrm{E}(S, r) \}.
$$
It is easy to show that barycenter (mean) of $S$ minimizes the function $\mathrm{E}(S,\cdot)$ (for more information see \cite{Ding,Fisher}). The above consideration can be precisely formulated as follows:
\begin{twr}[$k$-means] \label{theorem:3.1}
Let $S$ be a finite subset of $\mathbb{R}^{N}$. We have
$$
	\mathrm{E}(S,  \mu(S)) = \inf _{r \in \mathbb{R}^N} \{\mathrm{E}(S, r) \}
$$ 
where $ \mu( S ):=\frac{1}{ \mathrm{card} S }\sum_{ s \in S } s $ denotes the  barycenter of $S$.
\end{twr}
Thus in the $k$-means the goal is to find such clustering $S=S_1\cup\ldots\cup S_k$ that the function
$$
	\mathrm{E} (  S_1,\ldots,S_k) =  \sum_{j=1}^{k} \mathrm{E}(S_{j},\mu(S_{j}))
$$
is minimal. $k$-means algorithm for the set $S$ after S. Lloyd \cite{Forgey,Lloyd,Mac} proceeds as follows:

\begin{center}
		\begin{algorithmic}
			\STATE {\bf stop condition}
			\STATE\hspace\algorithmicindent \textit{choose} $\varepsilon > 0$			
			\STATE {\bf initial conditions}
			\STATE\hspace\algorithmicindent \textit{choose} randomly points $\{ \overline{s}_1, \ldots, \overline{s}_k \} \subset S$
			\STATE\hspace\algorithmicindent \textit{obtain} first clustering ($S_{1}, \ldots, S_{k}$) by matching each of   
			\STATE\hspace\algorithmicindent the point $s \in S$ to the cluster $S_j$ specified by $\overline{s}_j$ such 
			\STATE\hspace\algorithmicindent that  $\| s - \overline{s}_j \|^{2}$ is minimal
			\REPEAT 
			\STATE\hspace\algorithmicindent let $\mathrm{E} = \mathrm{E}(S_1,\ldots,S_k)$
			\STATE\hspace\algorithmicindent \textit{compute} new points $\overline{s}_1, \ldots, \overline{s}_k$ which best ''describe"  
			\STATE\hspace\algorithmicindent  the clusters ($\overline{s}_{j} = \mu(S_{j}) $ for $j=1 , \ldots , k$) 
			\STATE\hspace\algorithmicindent \textit{obtain} new clustering ($S_{1} , \ldots , S_{k}$) by adding each of  
			\STATE\hspace\algorithmicindent the points  $s \in S$ to the cluster such that  $\| s - \overline{s}_j \|^{2}$ is 	
			\STATE\hspace\algorithmicindent minimal
			\UNTIL{$\mathrm{E}-\mathrm{E}(S_1,\ldots,S_k)<\varepsilon$}
		\end{algorithmic}
\end{center}

Lloyd's method guarantees a decrease in each iteration but does not guarantee that the result will be optimal.

\subsection{$(\omega,k)$-means}

In this chapter we consider generalization of $k$-means similar to that from the previous section concerning the Voronoi diagram. 
Instead of looking for the points which best ''describe'' clusters we seek $n$ dimensional subspaces of $\mathbb{R}^N$.

Let $S \subset \mathbb{R}^N$ and $\omega \in [0,1]^{n+1}$, $\sum \omega_j=1$ be fixed. For $\mathrm{v} \in \mathrm{E}_{n}(\mathbb{R}^{N})$ let
$$
	\mathrm{E} _{\omega}( S , \mathrm{v}  ) := \sum \limits_{s \in S } \mathrm{DIST}_{\omega}^{2} (s ,\mathrm{v} ).
$$
We interpret the function $\mathrm{E} _{\omega}( S , \mathrm{v}  )$ as an energy of the set $S$ respectively to the subspace generated by $ \mathrm{v}$. If the energy is zero, the set $S$ is subset of affine space generated by $\mathrm{v}$.
We say that $ \overline{ \mathrm{v} } $ best ''describes" the set $S$ if the energy  is minimal, more precisely if
$$
	\mathrm{E}_{\omega}(S,\overline{ \mathrm{v} }) = \inf_{\mathrm{v} \in \mathrm{E}_n(\mathbb{R}^N)} \{ \mathrm{E}_{\omega}(S,\mathrm{v}) \}.
$$
To obtain an optimal base we use a classical Karhunen-Loe\'ve transform (called also Principal Component Analysis, shortly PCA), see \cite{Jol}. 
The basic idea behind the PCA is to find the coordinate system in which the first few coordinates give us a "largest" possible information about our data. 

\begin{twr}[PCA]\label{theorem:3.2}
Let $S=\{s_1,\ldots,s_m\}$ be a finite subset of $\mathbb{R}^N$. Let 
$$
\mathcal{M}( S ) := (\mathrm{v}_{0} , \ldots,\mathrm{v}_{N}) \in \mathrm{E}_{N} (\mathbb{R}^N)
$$ 
be such that 
\begin{itemize}
	\item $\mathrm{v}_{0} = \mu (S)$;
	\item  $\mathrm{v}_1, \ldots,\mathrm{v}_N$ are pairwise orthogonal eigenvectors of $ [s_{1}-\mathrm{v}_{0}, \ldots, s_{m}-\mathrm{v}_{0} ] \cdot [s_{1}-\mathrm{v}_{0}, \ldots ,s_{m}-\mathrm{v}_{0} ]^{T}$ arranged in descending order (according to the eigenvalues)\footnote{$[s_{1}-v_{0} ,\ldots, s_{m}-\mathrm{v}_{0} ]$ is a matrix with columns $s_{j}-\mathrm{v}_{0}$, for $j=1 , \ldots , m $.}.
\end{itemize}
For every $ n < N $ and $\omega \in [0,1]^{n+1}$ we have
$$
	\mathrm{E}_{\omega}(S,\mathcal{M}_k(S)) = \inf_{\mathrm{v} \in \mathrm{E}_n(\mathbb{R}^N)} \{ \mathrm{E}_{\omega}(S,\mathrm{v}) \},
$$
where $\mathcal{M}_k(S):= (\mathrm{v}_{0}, \ldots , \mathrm{v}_{k})$.
\end{twr}
Thus given $\omega \in [0,1]^{n+1} $, $\sum \omega_j = 1$, in $(w,k)$-means our goal is to find such clustering $S=S_1\cup\ldots\cup S_k$ that the function
\begin{equation}\label{equ:four}
	\mathrm{E}_{\omega}( S_1,\ldots, S_k) := \sum_{j=1}^{k}  \mathrm{E}_{\omega}(S_{j},\mathcal{M}_n(S))
\end{equation} 
is minimal. Consequently $(\omega,k)$-means algorithm can be described as follows:
\begin{center}
		\begin{algorithmic}
			\STATE {\bf stop condition}
			\STATE\hspace\algorithmicindent \textit{choose} $\varepsilon > 0$			
			\STATE {\bf initial conditions}
			\STATE\hspace\algorithmicindent \textit{choose} randomly points $\{ \overline{s}_1, \ldots, \overline{s}_k \} \subset S$
			\STATE\hspace\algorithmicindent \textit{obtain} first clustering ($S_{1}, \ldots, S_{k}$) by matching each of  
			\STATE\hspace\algorithmicindent  the points $s \in S$ to the cluster such that  $\| s - \overline{s}_j \|^{2}$ is 
			\STATE\hspace\algorithmicindent minimal
			\REPEAT 			
			\STATE\hspace\algorithmicindent let $\mathrm{E} = \mathrm{E}_{\omega}(S_1,\ldots,S_k)$
			\STATE\hspace\algorithmicindent \textit{compute} vectors $ \mathrm{v}_1 , \ldots , \mathrm{v}_k $, which best ''describe" the 
			\STATE\hspace\algorithmicindent clusters, by the PCA method ($\mathrm{v}_{j} =\mathcal{M}_n(S_j) $) 
			\STATE\hspace\algorithmicindent \textit{obtain} new clustering ($S_{1}, \ldots, S_{k}$) by adding each of 
			\STATE\hspace\algorithmicindent the point $s \in S$ to the cluster such that  $\mathrm{DIST}_{\omega}(s,\mathrm{v}_{j})$
			\STATE\hspace\algorithmicindent  is minimal
			\UNTIL{$\mathrm{E}-\mathrm{E}_{\omega}(S_1,\ldots,S_k)<\varepsilon$}
		\end{algorithmic}
\end{center}

As is the case in the classical $k$-means, our algorithm guarantees a decrease in each iteration but does not guarantee that the result will be optimal  (cf. Example \ref{ex:kk}).
\begin{prz}\label{ex:kk}
As already mentioned in Section 2, the $k$-means do not find a global minimum and strongly depends on initial selection of clusters. In our case, this effect can be even more visible. Consider the case of circle $C$ in $ \mathbb{R}^{2} $ with 4 clusters and $ \omega = (0,1)$.
\begin{figure}[!t]
\centering
	\subfigure[local minimum]{\label{fig:cyc1}\includegraphics[width=1.5in]{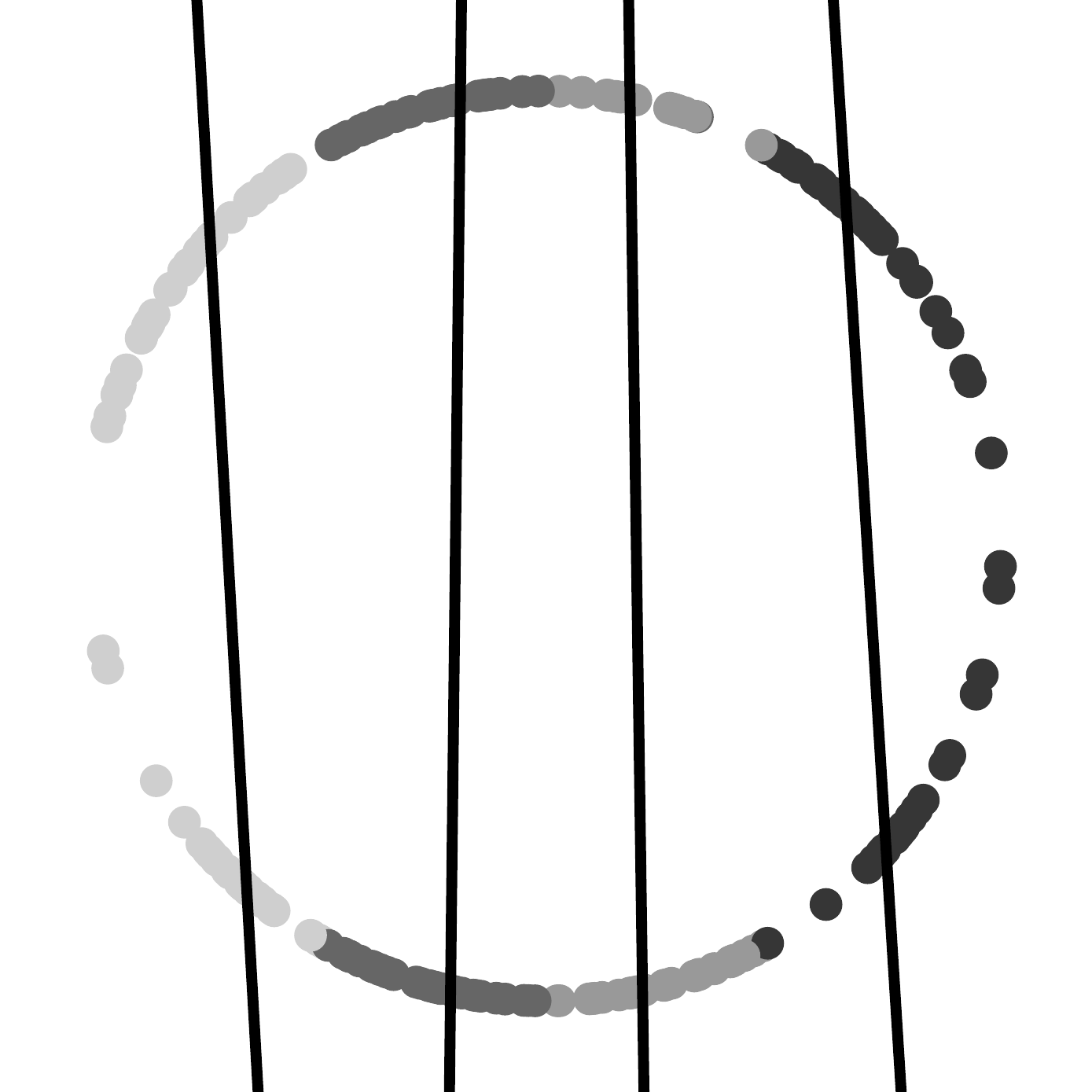}}\quad
	\subfigure[global minimum]{\label{fig:cyc2}\includegraphics[width=1.5in]{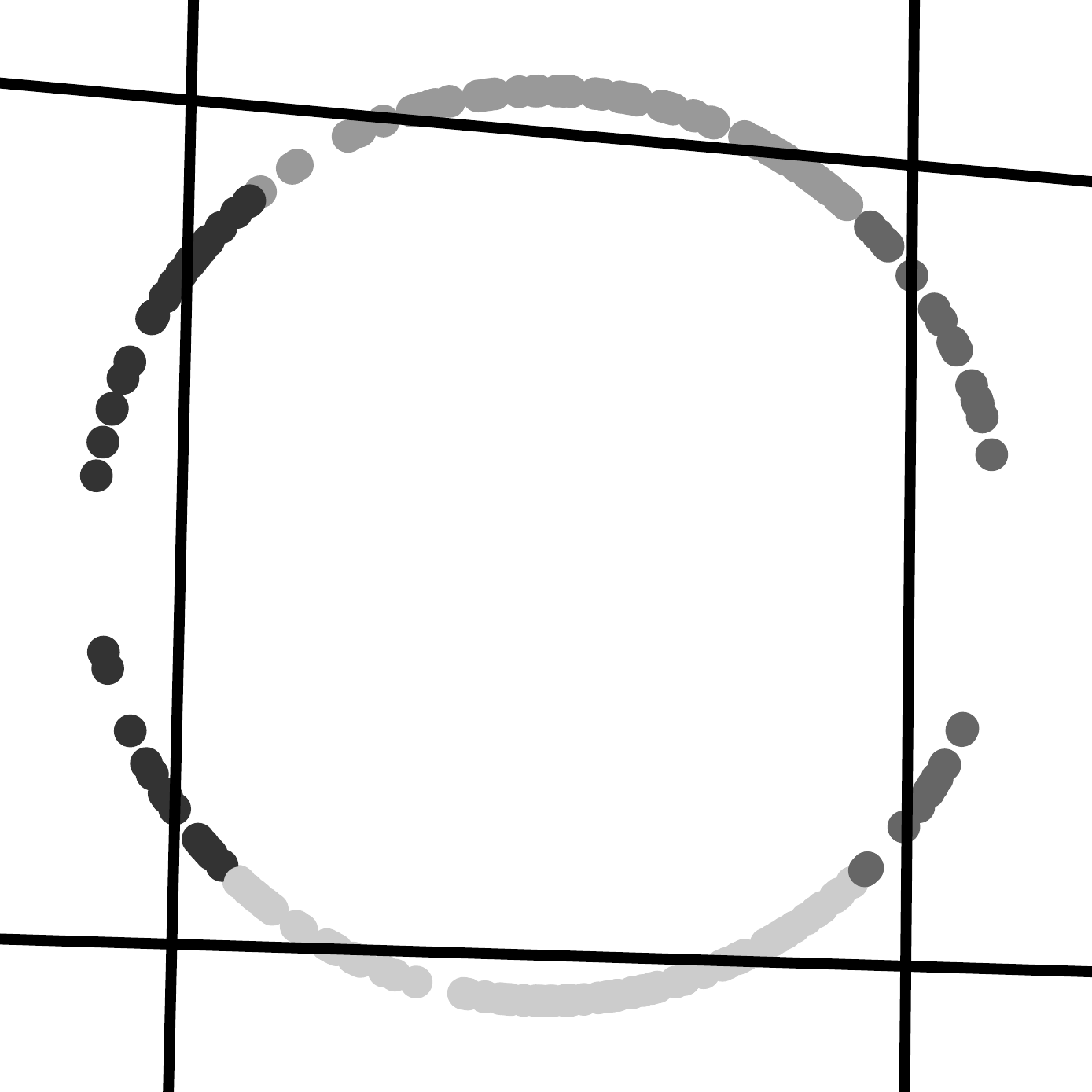}} 

	\caption{Circle clustering in $\mathbb{R}^2$ for 4 clusters with $\omega=(0,1)$. $(\omega,k)$-means method strongly dependents on initial conditions.}
	\label{fig:cyc_all}
\end{figure}
The picture, see Figure \ref{fig:cyc1}, shows clustering obtained by use $(\omega,k)$-means algorithm. Of course it is a local minimum of $\mathrm{E}^C_{\omega}$, however as we see at Figure \ref{fig:cyc2} it is far from being the global minimum.
\end{prz}

Initial cluster selection in our algorithm is the same as in $k$-mean algorithm, but it is possible to consider others ways:
\begin{itemize}
\item $k$-means++ algorithm \cite{kmeanspp};
\item starting from a given division (not from random distribution);
\item repeating the initial choice of clusters many times.
\end{itemize}
Each of above approaches usually solves the problem described in Example \ref{ex:kk}.

\begin{uw}
Let $ S \subset \mathbb{R}^{N} $ and $ \mathrm{v} \in \mathrm{E}_{n}(\mathbb{R}^N)$.
It is easy to notice that the above method has following properties:
\begin{itemize}
	\item for $\omega= (1 , 0 , \ldots , 0)$ we obtain the classical $k$-means, 
	\item for $n = 1$ we get Karhunen-Lo\'eve transform.
\end{itemize}
\end{uw}

As an algorithm's outcome we get:
\begin{itemize}
	\item division of the data into clusters $ \{S_1,\ldots,S_k  \} $;
	\item for each cluster an affine space of dimension $n$ obtained by the Karhunen-Lo\'eve method which best represents the given cluster.
\end{itemize}

\begin{prz}
If we apply our algorithm for regular plane subset (ex. square) we obtain generalized Voronoi diagram (cf. Fig. \ref{fig:wk_all}) -- Figure \ref{fig:wk_all_nn} present clustering for different weight vector changing from $\omega=(1,0)$ to $\omega=(0,1)$.

\begin{figure}[!t]
  \begin{center}
	\subfigure[]{\label{fig:wk1}\includegraphics[width=1in]{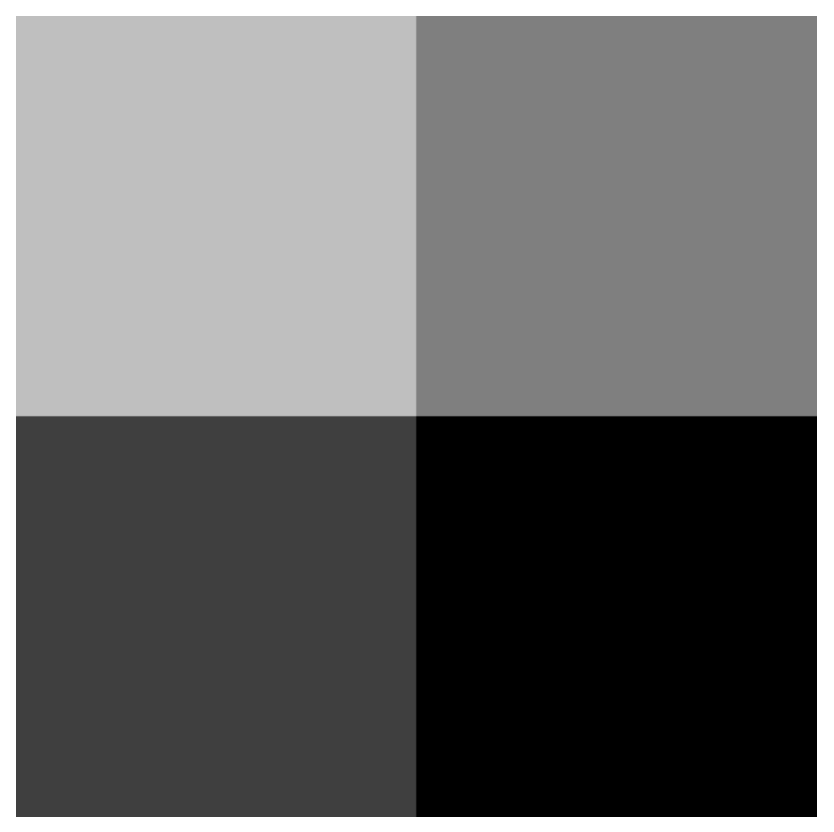}}\quad 
	\subfigure[]{\label{fig:wk1a}\includegraphics[width=1in]{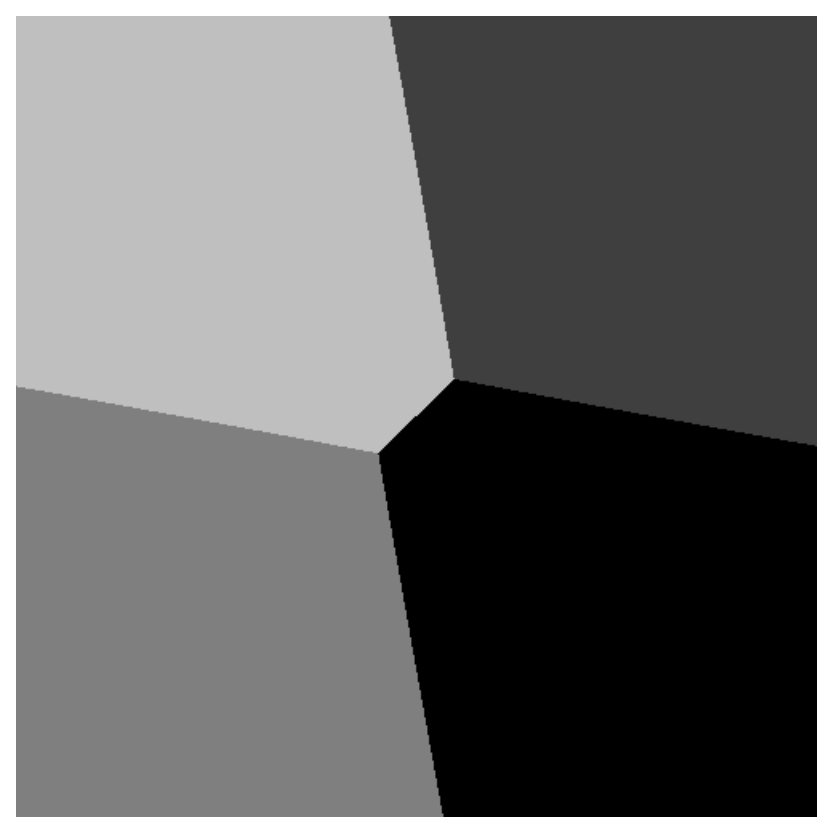}}\quad 
	\subfigure[]{\label{fig:wk2}\includegraphics[width=1in]{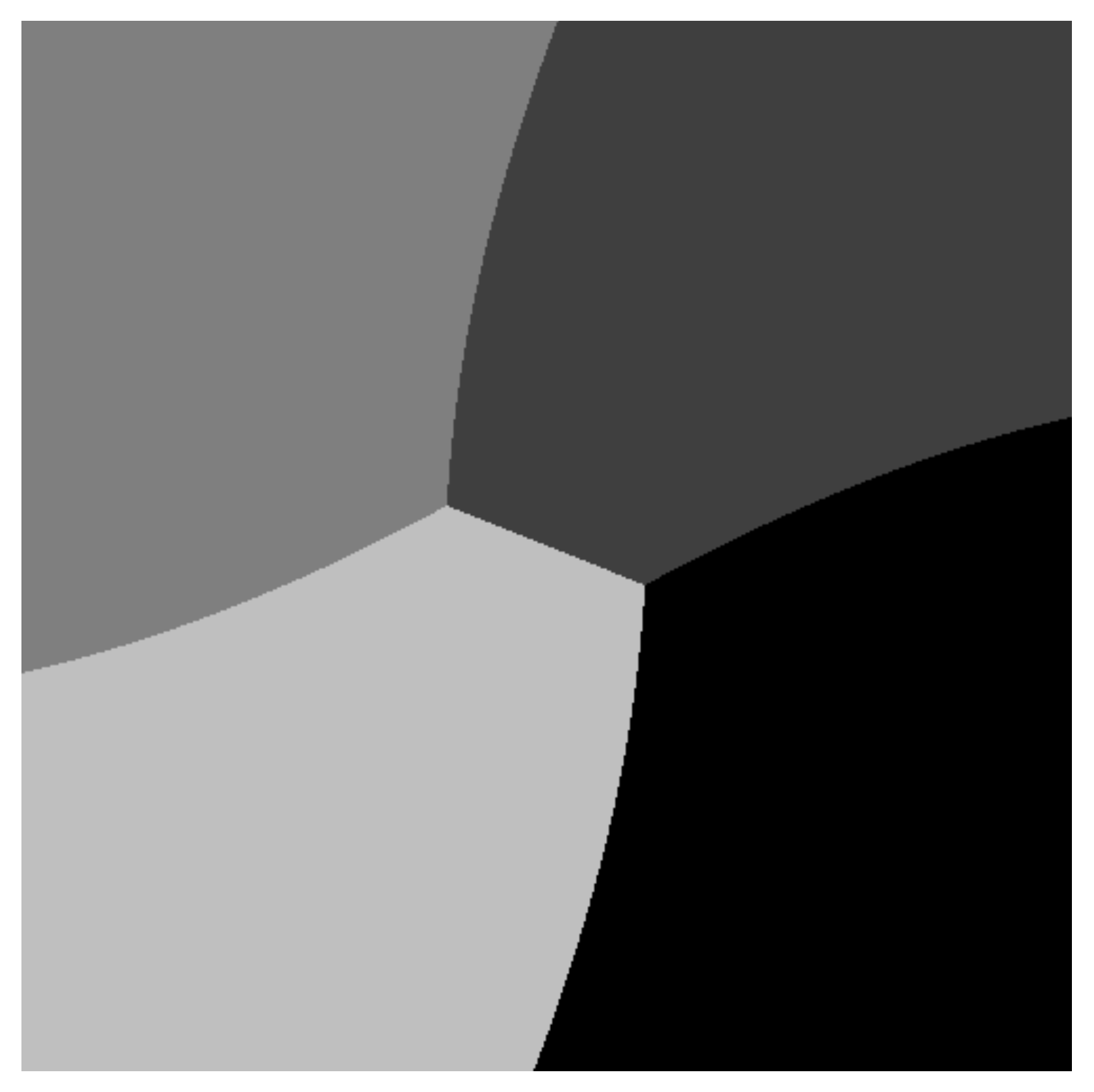}}\quad
	\subfigure[]{\label{fig:wk2a}\includegraphics[width=1in]{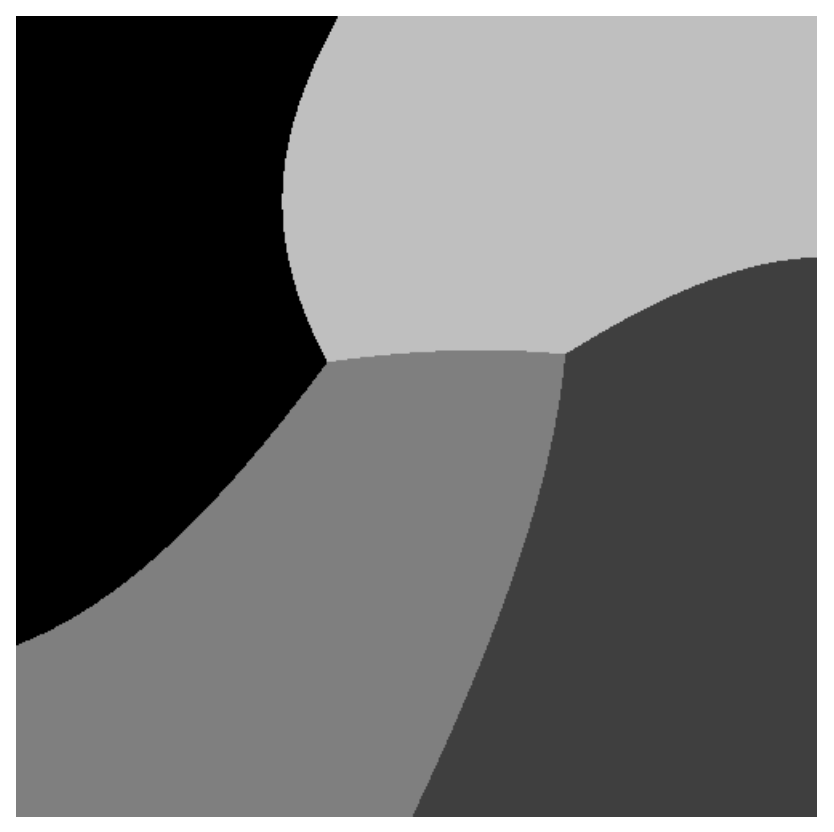}}\quad
	\subfigure[]{\label{fig:wk3}\includegraphics[width=1in]{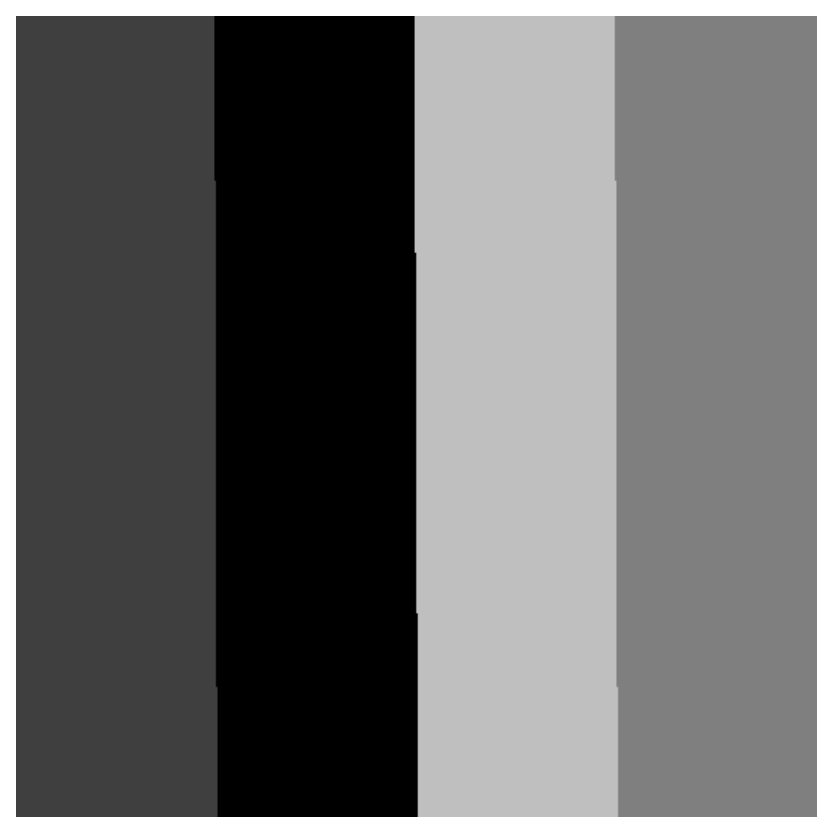}}\quad 
  \end{center}
	\caption{$(\omega,k)$-means method for clustering into 4 clusters of set $\{\frac{0}{1000},\frac{1}{1000},\ldots,\frac{1000}{1000}\}\times \{\frac{0}{1000},\frac{1}{1000},\ldots,\frac{1000}{1000}\}$ for different weight vectors: Fig. \ref{fig:wk000}, $\omega=(1,0)$; Fig. \ref{fig:wk025}, $\omega=(\frac 3 4,\frac 1 4)$; Fig. \ref{fig:wk050}, $\omega=(\frac 1 2,\frac 1 2)$; Fig. \ref{fig:wk075}, $\omega=(\frac 1 4, \frac 3 4)$; Fig. \ref{fig:wk100}, $\omega=(0,1)$.}
	\label{fig:wk_all_nn}
\end{figure}
\end{prz}

\section{Applications}

\subsection{Clustering}

\begin{figure}[!t]
	\centering
		\subfigure[]{\label{fig:patykc}\includegraphics[width=1.5in]{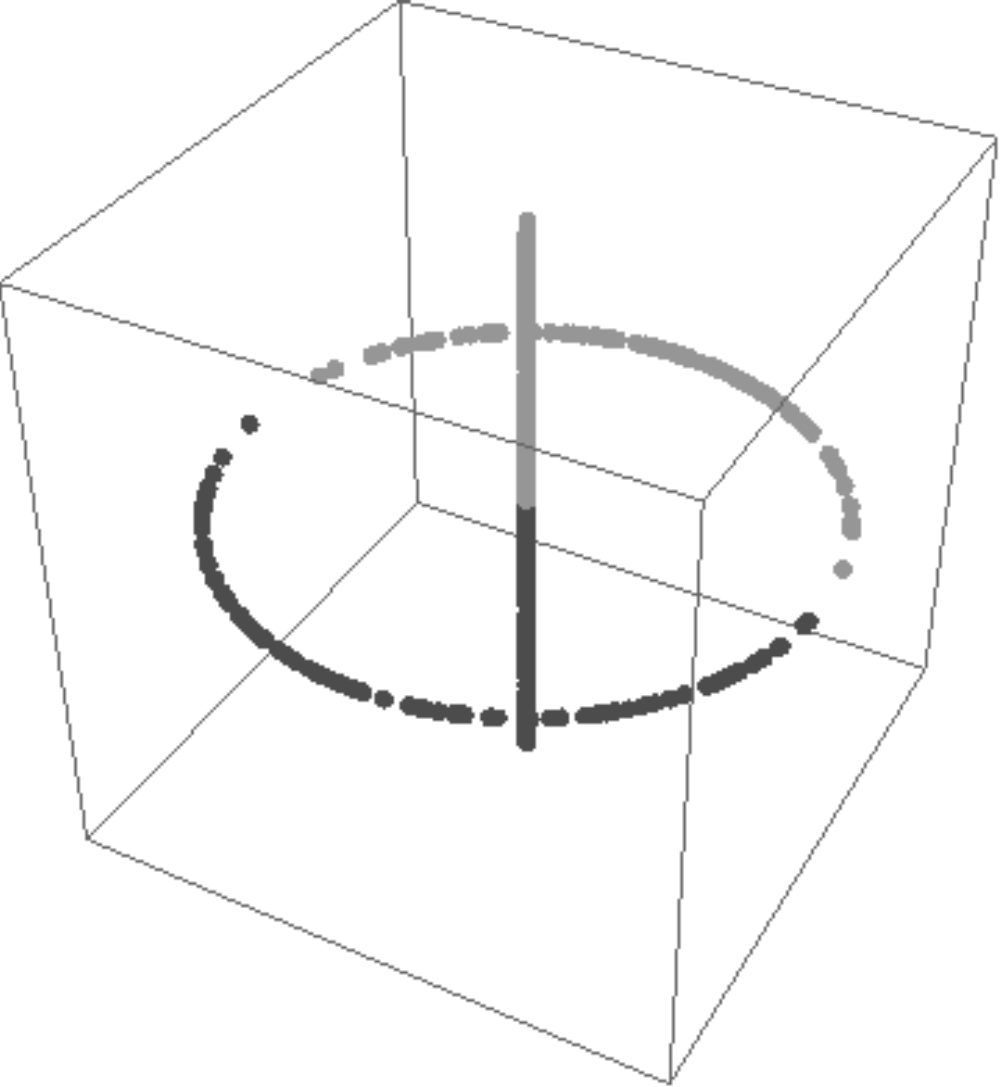}} \quad 
		\subfigure[]{\label{fig:patykn}\includegraphics[width=1.5in]{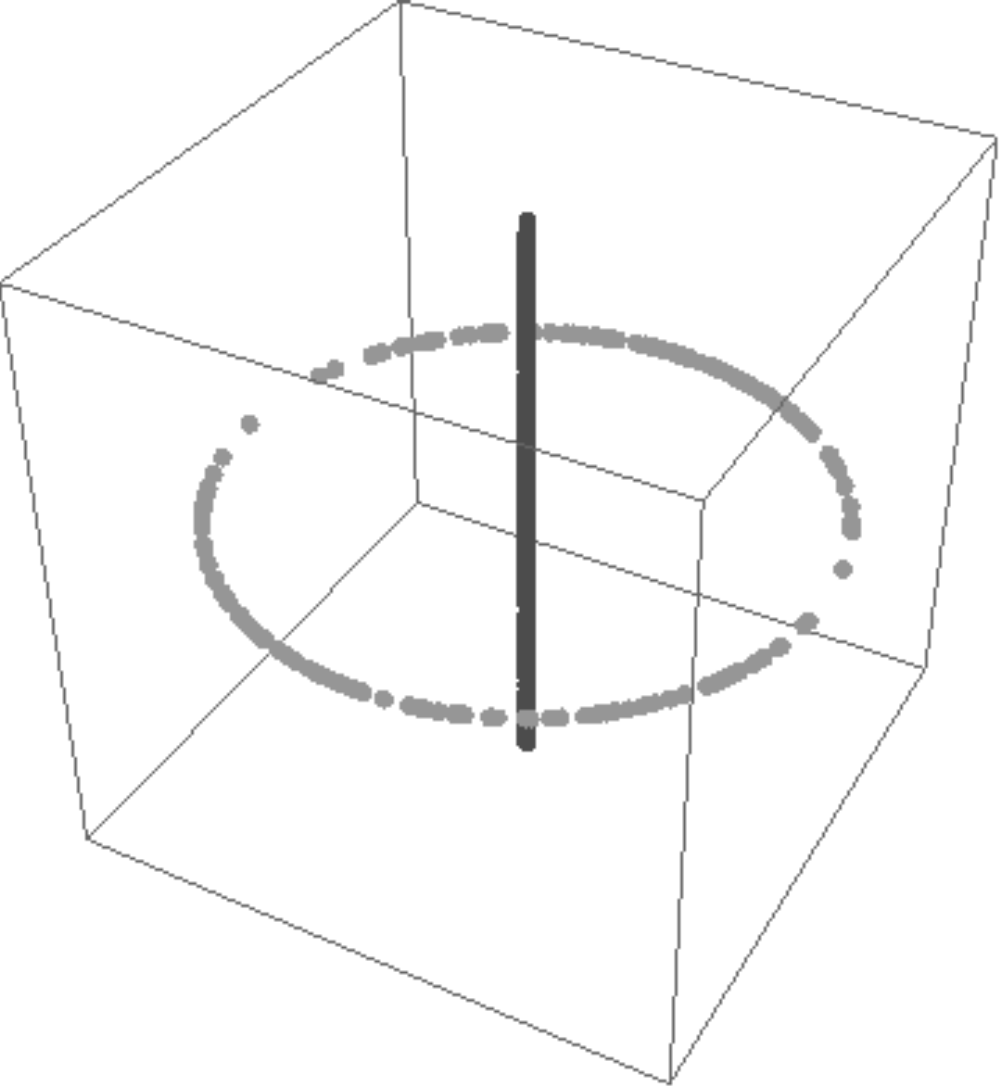}}
	\caption{Clustering with: Fig. \ref{fig:patykc} -- $k$-means; Fig. \ref{fig:patykn} -- $(\omega,k)$-means.}
	\label{fig:patnasz}
\end{figure}

Clustering, by $(\omega,k)$-means algorithm, gives a better description of the 
internal geometry of a set, in particular it found a reasonable
splitting into connected components of consider the points grouped along two parallel sections (see Figure \ref{fig:klus-line}).
Similar effect we can see in next example, when we consider the points grouped along circle and interval, see Figure \ref{fig:patnasz}. 

Concluding, in many cases the $(\omega,k)$-means method can be very useful in seeking $n$-dimensional (connected) components of given data sets.

\subsection{Analysis of Functions}

In this subsection we consider real data from acoustics. Acoustical engineers \cite{pilch} study reverberation which is observed when a sound is produced in an enclosed space causing a large number of echoes to build up and then slowly decay as the sound is absorbed by the walls and the air. Reverberation time is crucial for describing the acoustic quality of a room or space. It is the most important parameter for describing sound levels, speech intelligibility and the perception of music and is used to correct or normalize building acoustics and sound power measurements. 

\begin{figure}[!t]
  \centering
	\subfigure[]{\label{fig:vibro_signal}\includegraphics[width=0.45\textwidth]{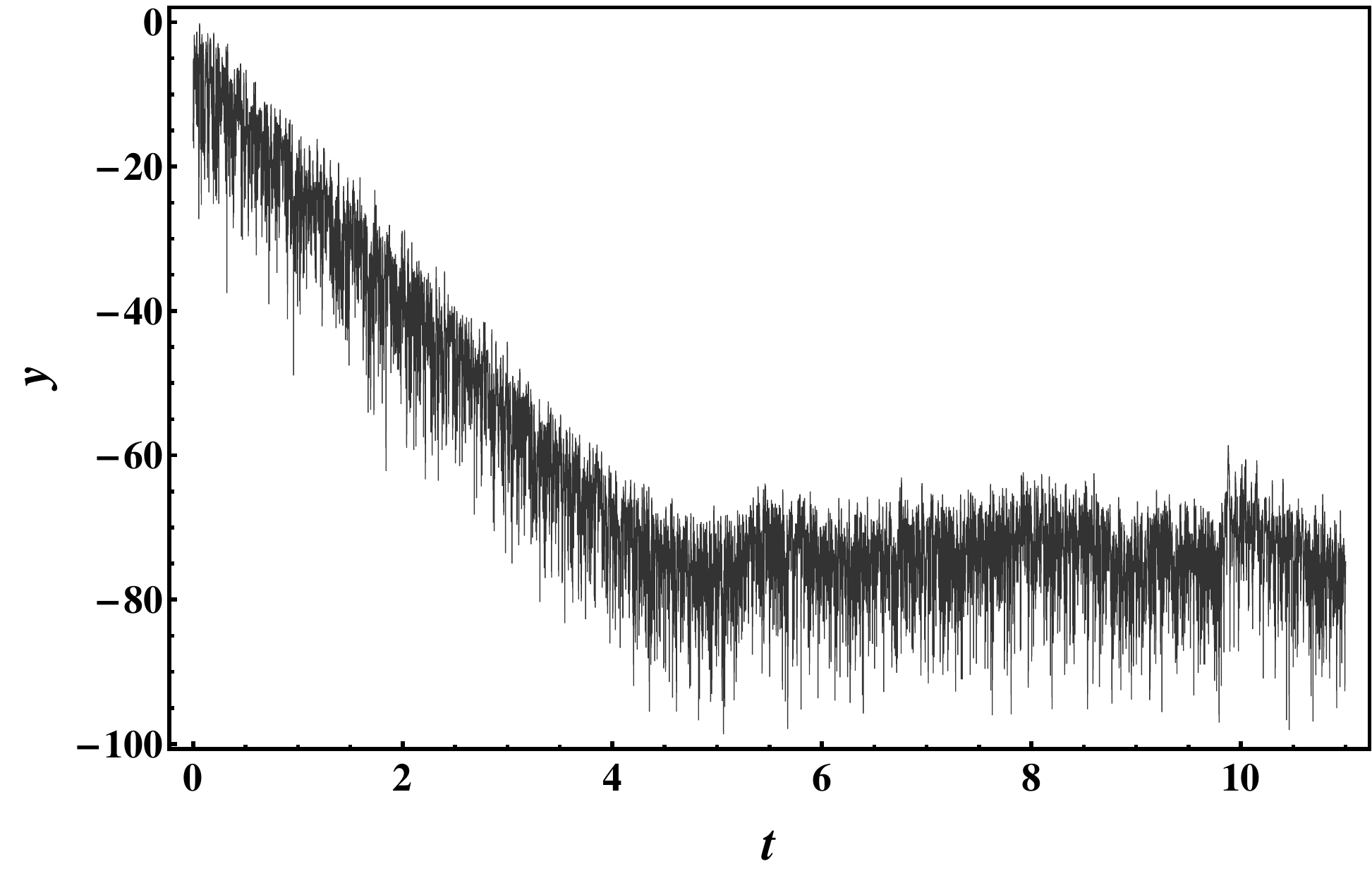}}\\ 
	\subfigure[]{\label{fig:vibro_clus}\includegraphics[width=0.45\textwidth]{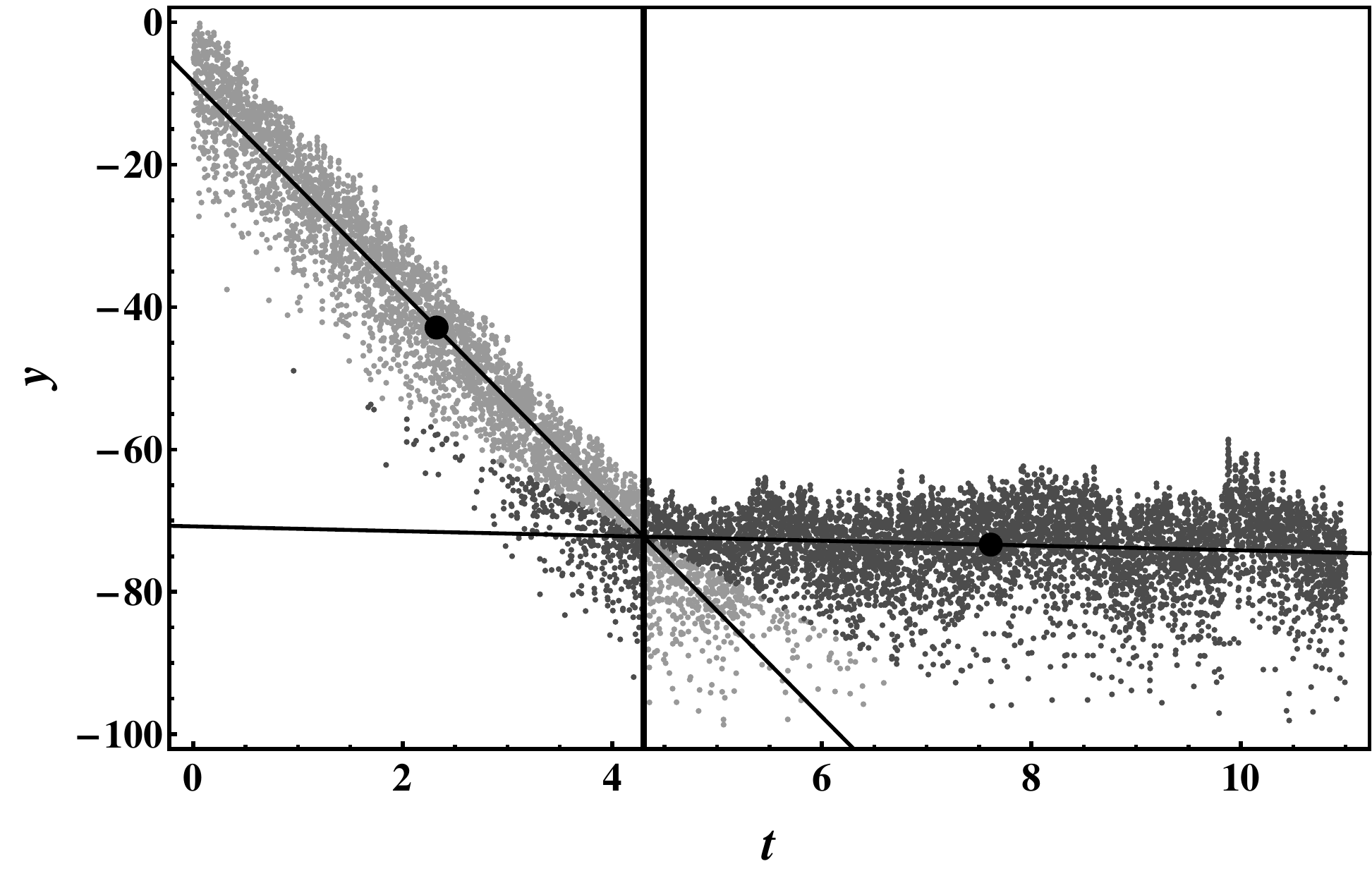}}
	\caption{Linear component of the data structure. Fig. \ref{fig:vibro_signal} -- decay curve (original data). Fig. \ref{fig:vibro_clus} -- outcome from $(\omega,k)$-means algorithm for $k=2$, $\omega=(0,1)$ we extract two linear components in data (black dots match clusters centers with the corresponding lines describing those clusters, vertical line separate sound and background noise -- after $4.3$ s).}
	\label{fig:co_all}
\end{figure}

We analyze the decay curve (see Figure \ref{fig:vibro_signal}) which presents measurement of sound level in time and describe way in which sound impulse vanishing into background noise. Based on this we want to recover reverberation time. In particular we know that we have two linear component: first connected with sound absorption by the space and second -- background noise. To use statistical analysis, we have to extract both of them. Our algorithm detects the $n$-dimensional subspaces, so for the parameters $\omega = (0,1)$ we obtain linear approximation of function, see Figure \ref{fig:vibro_clus}.

Results obtained using our algorithm are comparable with those obtained by classical methods and give more opportunities for further research.

\subsection{Image compression}

Our algorithm can be used to compress images. First, we interpret photo as a matrix. We do this by dividing it into 8 by 8 pixels, where each pixel is described (in RGB) by using 3 parameters. Each of the pieces is presented as a vector from $\mathbb{R}^{192}$. By this operation we obtain dataset from $ \mathbb{R}^{192} $. 

\begin{figure}[!t]
  \centering
	\subfigure[$k$-means: $k=5$, $n=0$]{\label{fig:lena_a}\includegraphics[width=1.5in]{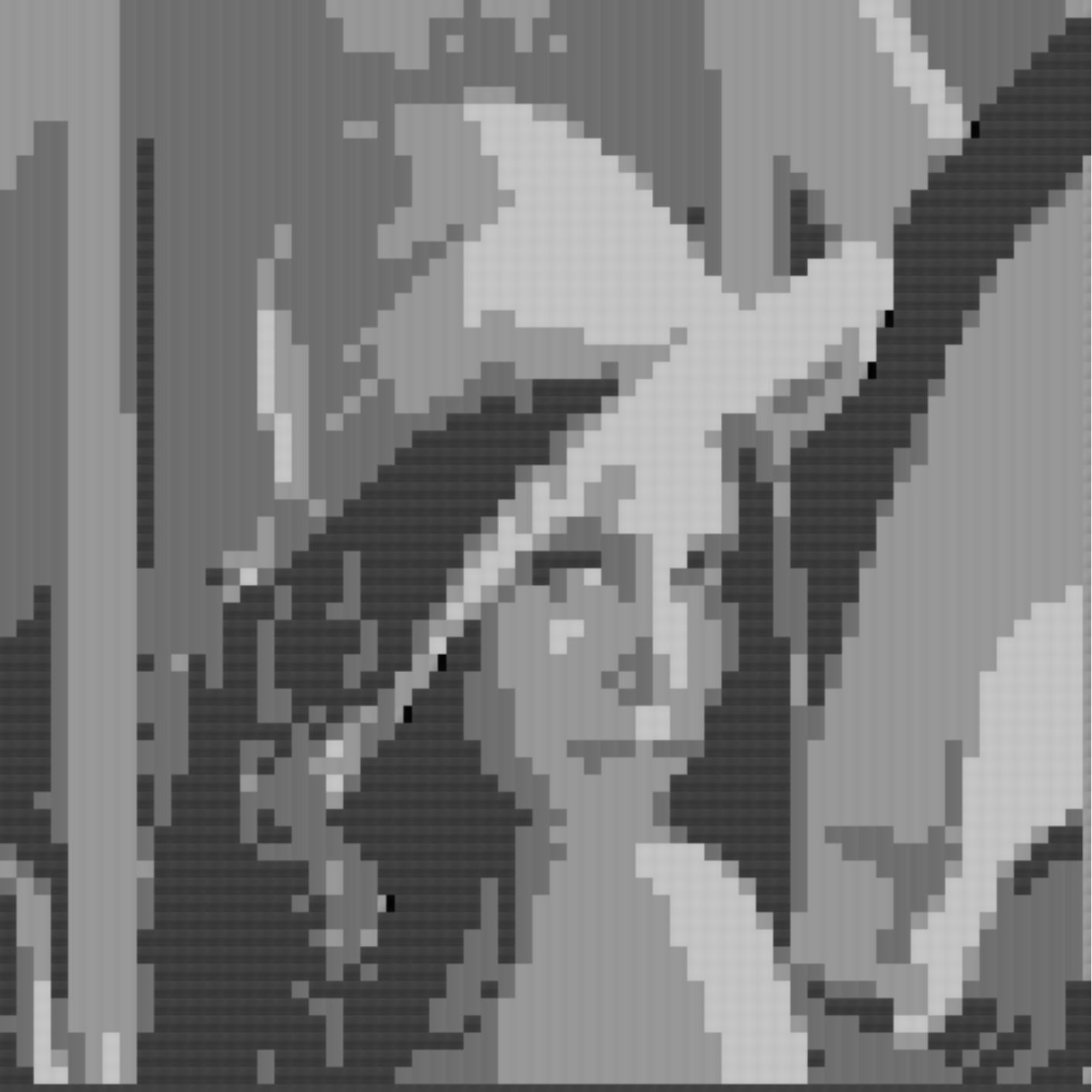}}\quad  \quad  
	\subfigure[PCA: $k=1$, $n=1$]{\label{fig:lena_b}\includegraphics[width=1.5in]{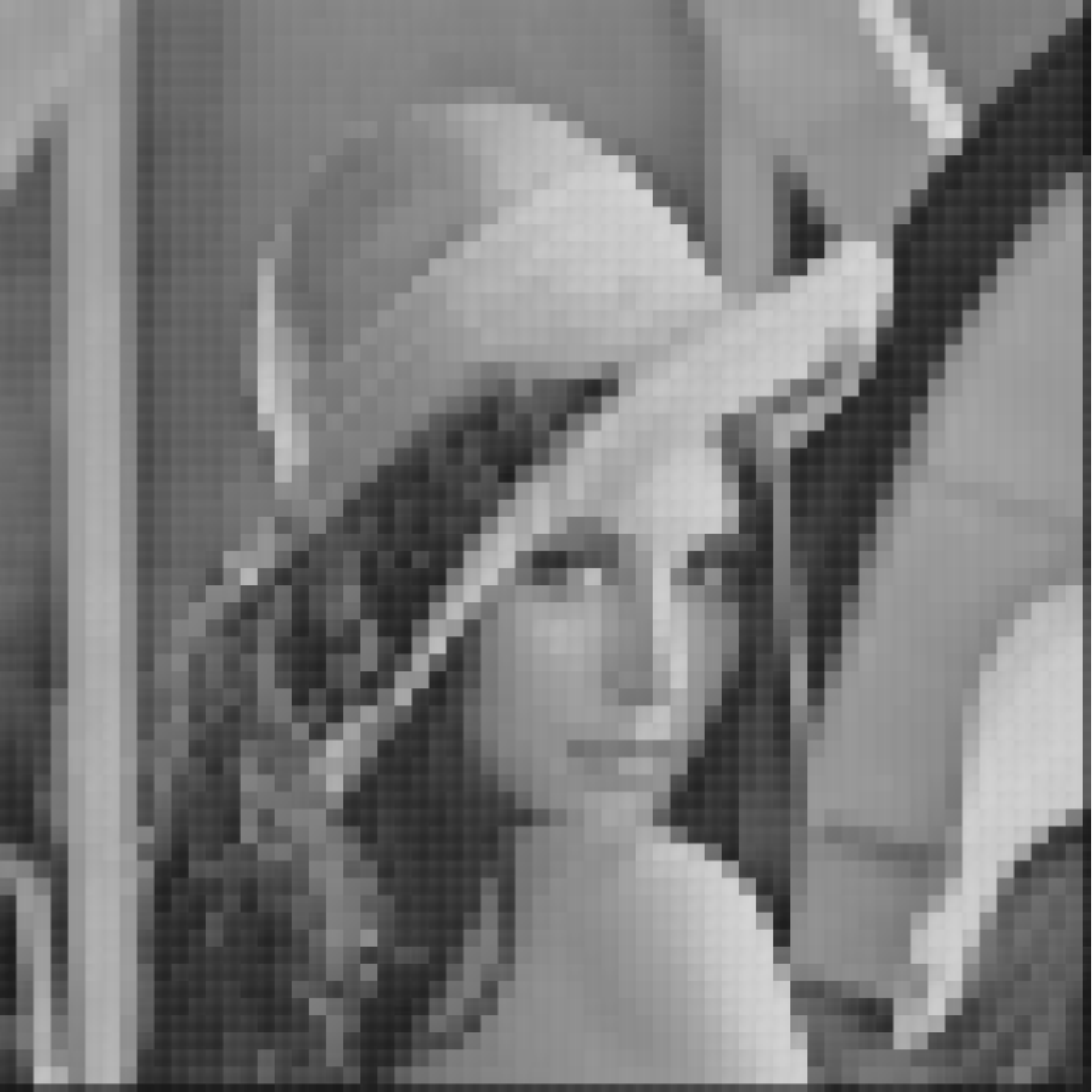}} \\
	\subfigure[$(\omega,k)$-means: $k=5$, $n=1$]{\label{fig:lena_c}\includegraphics[width=1.5in]{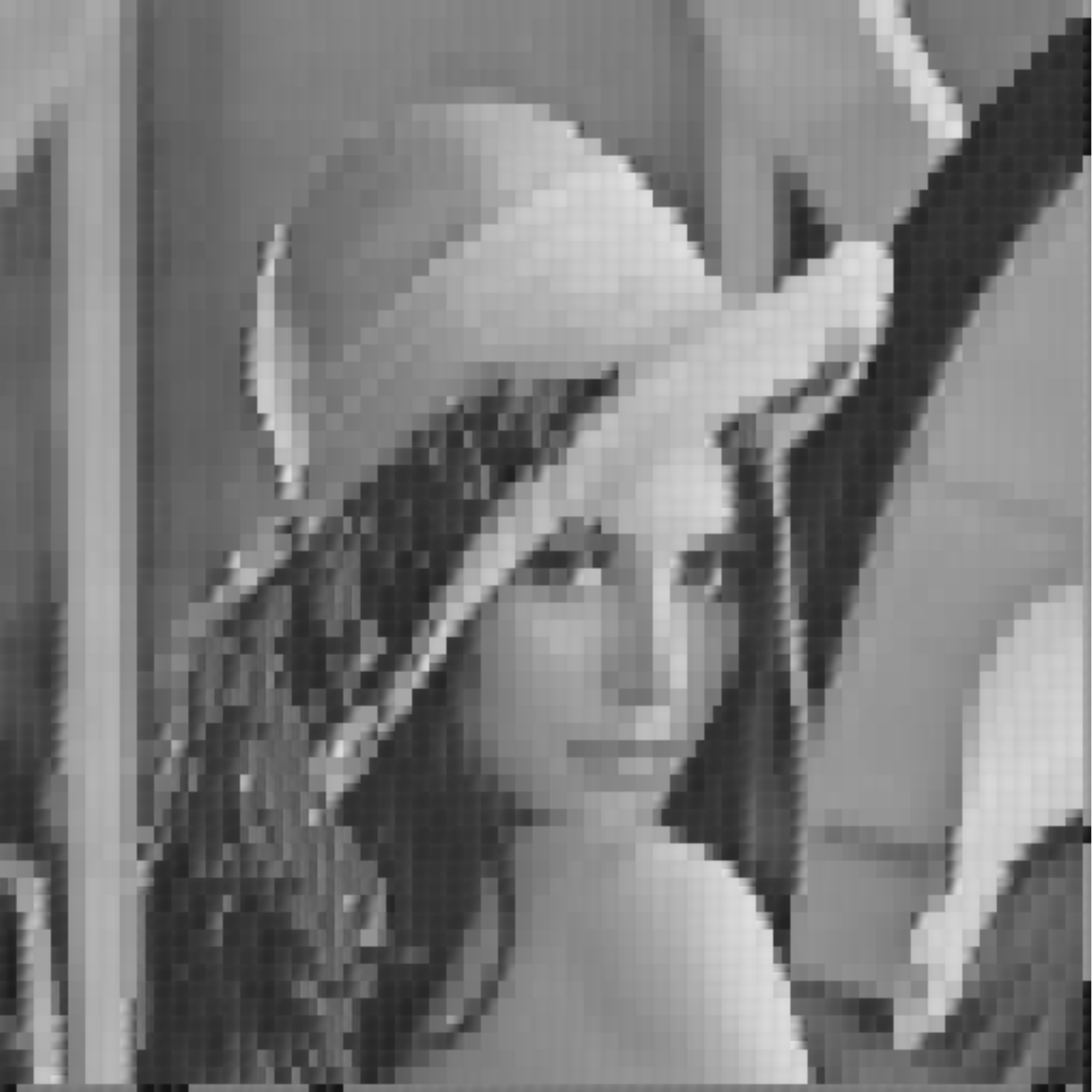}}\quad \quad 
	\subfigure[$(\omega,k)$-means: $k=5$, $n=5$ ]{\label{fig:lena_d}\includegraphics[width=1.5in]{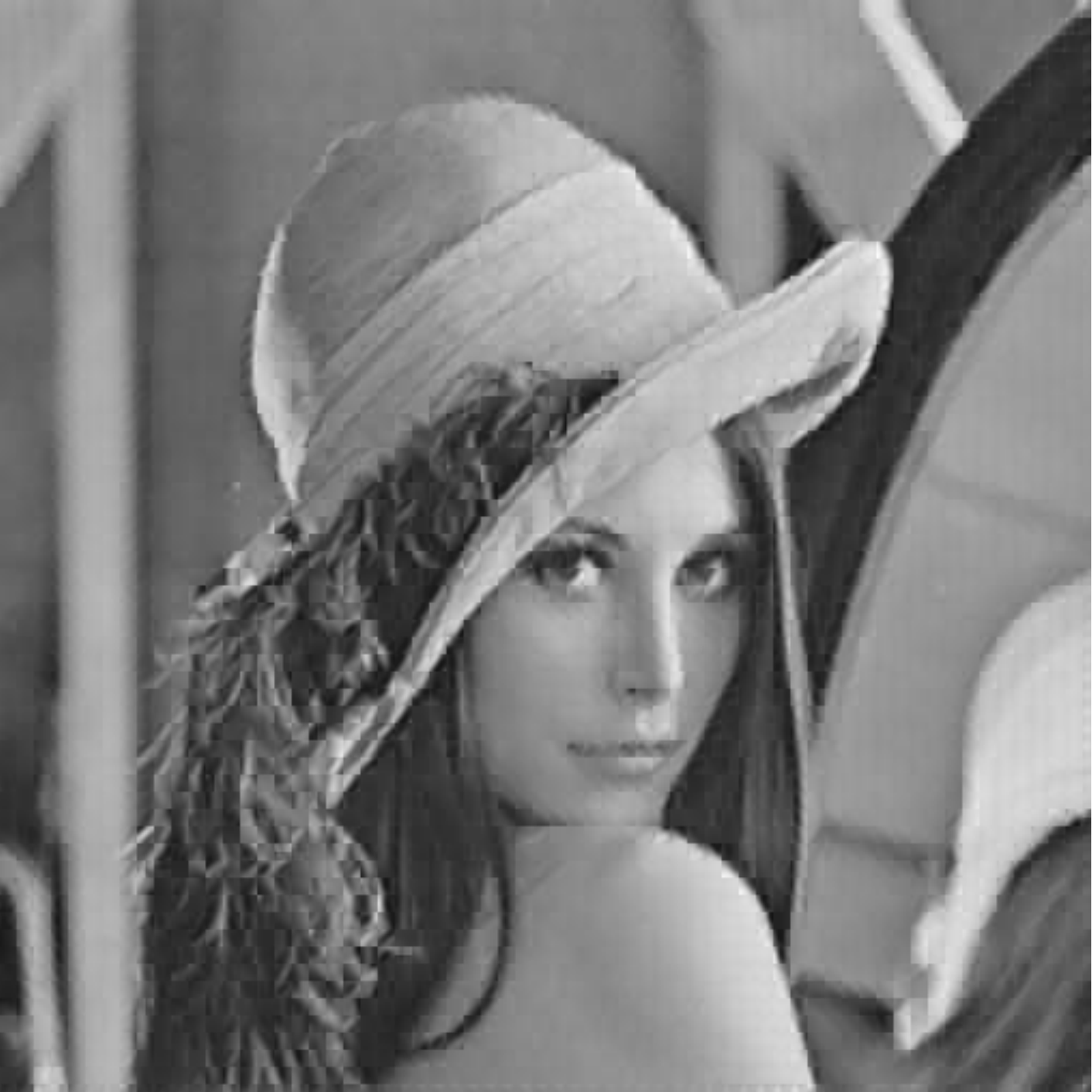}} 
	\caption{Compressed version of Lena picture. Subimage compare: Fig. \ref{fig:lena_a} -- classical $k$-means; Fig. \ref{fig:lena_b} -- Karhunen-Lo\'eve Transform; Fig. \ref{fig:lena_c} and Fig. \ref{fig:lena_d} -- $(\omega,k)$-means algorithm.}
\end{figure}

Taking into consideration the classical Lena picture ($508 \times 508$ pixels), let us present its compressed version with the use of $k$-means method (Figure \ref{fig:lena_a}, $k=5$, $n=0$), Karhunen-Lo\'eve Transform (Figure \ref{fig:lena_b}, $k=1$, $n=1$) and $(\omega,k)$-means algorithm (Figures \ref{fig:lena_c} and \ref{fig:lena_d}).
As we can see the algorithm allows to reconstruct with great accuracy compressed images while reducing the amount of needed information to save (in our example we remember ex. only 5 coordinates in 192-dimensional space).

Table \ref{tab:1} presents error in image reconstruction for Lena picture. We run $(\omega,k)$-means algorithm 16 times and each run improve clustering  quality 50 times. 
\begin{table}[!t]
\centering
\caption{Error in image decompression for certain $k$ and $n$.}
\label{tab:1}
\begin{IEEEeqnarraybox}[\IEEEeqnarraystrutmode\IEEEeqnarraystrutsizeadd{2pt}{0pt}]{x/r/Vx/r/v/r/v/r/v/r/v/r/v/r/x}
\IEEEeqnarraydblrulerowcut\\
&&&&\IEEEeqnarraymulticol{11}{t}{$n$}&\\
&\hfill\raisebox{-3pt}[0pt][0pt]{$k$}\hfill&&\IEEEeqnarraymulticol{13}{h}{}%
\IEEEeqnarraystrutsize{0pt}{0pt}\\
&&&&\hfill0\hfill&&\hfill1\hfill&&\hfill2\hfill&&\hfill3\hfill&&\hfill4\hfill&&\hfill5\hfill&\IEEEeqnarraystrutsizeadd{0pt}{2pt}\\
\IEEEeqnarraydblrulerowcut\\
&1&&&40328.6 && 19499.3 && 16358.1 && 12452.1 && 10160.5 && 8149.7&\\
&2&&&27502.8 && 17193.3 && 13031.9 && 10382.8 && 9082.7 && 7913.2&\\
&3&&&23261.2 && 15437.2 && 11631.8 && 9612.0   && 8350.1 && 7358.3&\\
&4&&&20990.4 && 14454.1 && 11004.1 && 9192.7   && 7922.4 && 7095.9&\\
&5&&&20150.1 && 13740.0 && 10602.4 && 8867.2   && 7745.9 && 6814.5&\\
\IEEEeqnarraydblrulerowcut\\
\end{IEEEeqnarraybox}
\end{table}

\section{Implementation}

Sample implementation of $(\omega,k)$-means algorithm prepared in Java programming language is available at \url{http://www.ii.uj.edu.pl/~misztalk}.

\ifCLASSOPTIONcaptionsoff
  \newpage
\fi



\bibliographystyle{IEEEtran}
\bibliography{IEEEabrv,../bib/paper}
\end{document}